\documentclass[iop]{emulateapj}
\usepackage{graphicx}
\usepackage{xspace}
\newcommand{\MS}{\mbox{M$_{\textstyle \odot}$}}

\newcommand{\RS}{\mbox{R$_{\textstyle \odot}$}}

\newcommand\nifsx{${}^{56}$Ni\xspace}
\newcommand\cofsx{${}^{56}$Co\xspace}
\newcommand\fefsx{${}^{56}$Fe\xspace}

\slugcomment{Astrophys.~J., 2016, in press}

\shorttitle{SLSN-I in Non-Hydrogen Envelopes}
\shortauthors{Sorokina et al.}

\begin{document}

\title{Type I Superluminous Supernovae\\ as Explosions inside Non-Hydrogen Circumstellar Envelopes}


   \author{Elena Sorokina\altaffilmark{1,2,3*}, Sergei Blinnikov\altaffilmark{2,3,4},
           Ken'ichi Nomoto\altaffilmark{3\dag}, Robert Quimby\altaffilmark{3,5}, Alexey Tolstov\altaffilmark{3}}


\affil{\altaffilmark{1}Sternberg Astronomical Institute, M.V.Lomonosov Moscow State University,
119991 Moscow, Russia}
\affil{\altaffilmark{2}Institute for Theoretical and Experimental Physics, 117218 Moscow, Russia}
\affil{\altaffilmark{3}Kavli Institute for the Physics and Mathematics of the Universe (WPI),
      The University of Tokyo Institutes for Advanced Study, The University of Tokyo, Kashiwa, Chiba 277-8583,
      Japan}
\affil{\altaffilmark{4}All-Russia Research Institute of Automatics (VNIIA), 127055 Moscow, Russia}
\affil{\altaffilmark{5}Department of Astronomy, San Diego State University, San Diego, CA 92182, USA}

\altaffiltext{$^{*}$}{elka.sorokina@gmail.com}
\altaffiltext{\dag}{Hamamatsu Professor}


%



\begin{abstract}
A number of Type I (hydrogenless) superluminous supernova (SLSN)
events have been discovered recently.
However, their nature remains debatable.
One of the most promising ideas is the shock-interaction mechanism, but only
simplified semi-analytical  models have been applied so far.
We simulate light curves for several Type~I SLSN (SLSN-I) models enshrouded
by dense, non-hydrogen circumstellar (CS) envelopes, using a
multi-group radiation hydrodynamics code
that predicts not only bolometric, but also multicolor light curves.
We demonstrate that the bulk of SLSNe-I including those with relatively
narrow light curves like SN~2010gx
or broad ones like PTF09cnd can be explained by the
interaction of the SN ejecta with the CS envelope,
though the range of parameters for these models is rather wide.
Moderate explosion energy ($\sim (2 - 4)\cdot 10^{51}$~ergs) is sufficient to
explain both narrow and broad SLSN~I light curves,
but  ejected mass and  envelope mass differ for those two cases.
Only 5 to 10~\MS\ of non-hydrogen material is needed to reproduce
the light curve of SN~2010gx,
while the best model for PTF09cnd is very massive:
it contains almost $ 50\, M_\odot $ in the CS envelope and only $ 5\, M_\odot $ in the ejecta.
The CS envelope for each case extends from 10\RS\ to $\sim 10^5$\RS\ ($7\cdot 10^{15} $~cm),
which is about an order of magnitude larger than typical photospheric radii of standard SNe near the maximum light.
We briefly discuss possible ways to form such unusual envelopes.

\end{abstract}

\keywords{ Circumstellar matter --- shock waves --- supernovae: general
               }


\section{INTRODUCTION}

Quite a number of observations of Type I and II superluminous supernovae (SLSNe, SLSNe-I, and SLSNe-II) have
appeared during the last few years
\citep{Ofek2007,Smith2007,Smith2010,GalMaz2009,Pastorello2010,Young2010,QuimbyNatur2011,Quimby2013}.
The definitions of SLSN-I, SLSN-II, and other relevant material are reviewed in~\citet{GalYam2012}.
There are signs that these objects have been observed even at high redshifts, $z=2-4$ \citep{CookeHighZ}.
The mechanism of the explosion for such
bright and long lasting events is still not fully understood. One can consider either an
unusually energetic explosion through which a huge amount of \nifsx is produced
\citep{NomTom2007,UmedaNom99as,Moriya2010,Langer12,KozBlLangYoon,YoshidaOkitaUmeda},
or an additional energy source, like a millisecond magnetar, that somehow transforms its rotation energy
into the energy of the SN ejecta \citep{KasenBildsten,Inserra13,NichollMagnetar},
or an interaction of the ejecta with an extended and dense circumstellar (CS) envelope,
which we focus on in this paper.

\citet{FalkArnett77} have shown that the width of the SN light curve depends
on the mass through which the shock wave breaks out.
If an exploding star possesses just a standard hydrostatic envelope,
like an atmosphere of a supergiant star,
the broadening of its light curve cannot be extremely large.
However, sometimes the star, especially the massive one, can be surrounded by rather dense and extended envelope
originated from slow wind, preexplosions, or stellar mergers.
In Section~\ref{sec:summary} we will discuss several possible ways to form such an envelope
in a little more detail.

The idea of producing large radiative flux during the interaction of the gas ejected in two subsequent explosions
was suggested by \citet{GraNad1986} for the explanation of SNe~IIn.
\citet{ChugaiEa04} successfully applied this model to the explanation
of spectral and light curve features of type IIn SN~1994W.
A physical mechanism for those multiple explosions (pulsational pair instability) was  proposed by \cite{HegWoo2002}.
\cite{WooBliHeg2007} used  this model to explain Type II superluminous
SN~2006gy as a moderately energetic explosion
($\sim 3\cdot 10^{51}$~ergs) without any radioactive material.

SN light curves for the interacting model have been constructed analytically in a number of papers.
\citet{ChevIrwin} compared the light curves for optically thin vs. optically thick cases.
\citet{Moriya13analit} found a good agreement of analytical interacting models with the light curves
of SLSNe~IIn.
\citet{Chatz13} present several semi-analytical models of the SLSN light curves for three
feasible power inputs mentioned above and their combinations.
They found that the ejecta-CSM interaction model provided a better fit to the light curves
of most of the observed hydrogen-rich events (SLSNe-II) they investigated,
and they suggested that the same mechanism can be applied to hydrogen-poor events (SLSNe-I).

Numerically, the interaction model for SLSNe is the most difficult for calculation
among all the possibilities under consideration.
The cause of difficulties is the complicated hydrodynamical structure of the model.
Most of the gas is gathered into a thin and dense layer behind the radiation dominated shock wave,
which can lead to thermal and hydrodynamical instabilities.
There are only a very few numerical codes in the world
that are able to cope with this problem.
A few papers have been published in which SLSNe-II were modeled \citep{MoriyaSLSNeII13,Moriya13_09gy,Whalen13}.
The shock interaction with CSM is generally considered the most probable explanation
for the high luminosity of SLSNe-II.
In this paper, we will focus on another type of SLSNe: hydrogen-poor SLSNe-I.

The work was started in \citet{BlSorCOshells_arXiv},
where we followed \citet{Fryer2010} who calculated light curves for the explosion
caused by a carbon-oxygen (CO) white dwarf merger.
After the explosion, the model is hydrodynamically similar to the interacting SLSN model:
the explosion also happens inside a dense (but probably a bit less extended) envelope.
In \citet{BlSorCOshells_arXiv}, we show that in principle superluminous SN~2010gx can also be explained
by the interaction model with moderate values for the explosion parameters.
After that work, we have improved our radiation hydrodynamics numerical code {\sc stella}
\citep{BliEas1998,BliRop2006} in order to make it more appropriate to solve the SLSN problem.
In the current paper, we demonstrate a few more carbon-oxygen and helium interacting models  for SN~2010gx and PTF09cnd,
which are among supernovae with the narrowest and broadest light curves, respectively, known for SLSNe-I.
Both models require moderate explosion energy and a reasonable radius of the extended envelope,
though the mass of carbon in the envelope must be quite large to reproduce the broad light curve.
We briefly discuss some possibilities to create such a system through stellar evolution in the last section of the paper.
If these systems really exist
this would allow us to claim that the bulk of SLSNe might have a similar origin
and that most of their radiation might be produced during an interaction of SN ejecta
with extended envelopes or detached shells.
The differences between SLSNe-I can be explained by different CS envelope structures.
This does not relate to some very exotic objects, like ASASSN--15lh \citep{asassn},
which stands apart from the bulk of SLSNe-I due to its extremely high luminosity,
and origin of which is still questionable.

The structure of the paper is as follows.
In section~\ref{sec:presn} we describe the models we use.
Section~\ref{sec:simul} explains how we have carried out simulations of SN light curves.
In sections~\ref{sec:hydroprofiles} and \ref{sec:LCsp} we present the results of our calculations
and compare them to the observed broad band light curves of SN~2010gx and PTF09cnd.
In section~\ref{sec:summary} we compare the advantages and limitations of the interaction mechanism
to other mechanisms, such as magnetar-powered SNe and pair instability SNe,
and some feasible ways these dense circumstellar
envelopes might form.

\section{PRESUPERNOVA MODELS}
\label{sec:presn}

\begin{figure}
\centering
\includegraphics[width=0.7\linewidth]{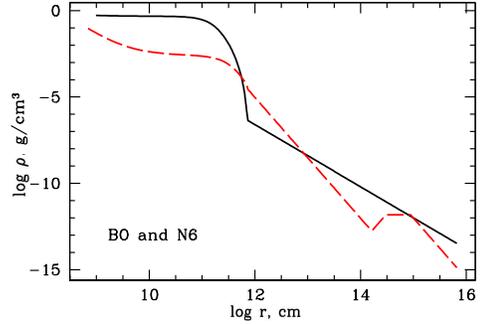}
\caption{\noindent Two typical examples of the initial density structures for our models.
The {\it solid} line shows the windy-like model B0 with the total mass 54 \MS.
The {\it dashed} line shows the model N6 (see Table~\ref{table:models}) with a detached shell;
the total mass of the model is 10 \MS.
The central part, which we call ``ejecta,'' has a polytropic structure for all initial models.
        }
\label{fig:init.density}
\end{figure}

All presupernova models for this work have been constructed
in the same way as we described in many papers \citep{ChugaiEa04,BaklEa05}.
A~quasi-polytropic structure in mechanical equilibrium is considered in the interior part,
which we  call ``ejecta''.
The temperature is related to the density as $T \propto \rho^{0.31}$.
This part has mass $M_{\rm ej}$ and radius $R_{\rm ej}$,
which is equal to 10~\RS\ for all models in this work.
$M_{\rm ej}$ can be much less than
the total mass of the collapsing core and the condition of the mechanical equilibrium
is not necessary; it is just a convenient form of parameterizing the models.

To make an interacting model, we surround the ejecta
with a rather dense   envelope with the mass $M_{\rm w}$ extended to the radius $R_{\rm w}$.
For all of our models, the outer radius of the CS envelope is about $10^5$~\RS, or $\sim 7\cdot 10^{15}$~cm.
For most of the models, the envelope adjoins the ejecta
without any jump in density between the ejecta and the envelope
and has a power-law density distribution $\rho \propto r^{-p}$,
which simulates the envelope or the wind that surrounds the exploding star.
We will refer to this structure as the ``extended envelope.''
For a steady wind, $p=2$,
but in the very last stages of the evolution of a presupernova star the wind may not be steady.
For our models, we vary $p$ in the range between 1.5 and~3.5.
We also tried  another kind of density distribution in the CS envelope:
we construct a model with the envelope  concentrated within a shell
detached from the ejecta by a region of lower density.
The density falls down and raises up still with power-law behavior,
but the slope of its distribution for this kind of model is steeper
than in the windy-like models we described above.
We will call this structure ``a detached shell.''
The density distributions for a couple of typical models are shown in Figure~\ref{fig:init.density}
in order to illustrate these two kinds of structures.
No attempt is done to keep equilibrium in the CS envelope, but its dynamical time scale
is so large that no appreciable motion has developed
during the time of the light curve simulation.

We calculate light curves for SNe exploding within these envelopes.
A shock wave forms at the border between the ejecta and the envelope.
It  very efficiently converts
the energy of the ordered motion of expanding gas to that of the chaotic
thermal motion of particles, which can be easily emitted.
As a result, we expect to
derive light curves bright enough to explain at least  part of
SLSNe-I without an assumption of unusually high explosion energy.

Chemical elements in all but one of our models are distributed uniformly.
Typically, we use CO models with different C to O ratios or helium models.
We always add some elements  with higher atomic numbers (usually, 2\% of the total mass)
with the abundances in solar proportion.
The models do not contain any radioactive elements
since the goal of the current work is to check the effect of pure ejecta-CSM interaction.
We will show the influence of radioactive \nifsx\ to the light curve in the next paper
(A.Tolstov et al. 2016, in preparation).

All models initially have  $T = 10^3$~K in the envelope.
Higher envelope temperatures produce a spurious flash of light emitted by the huge envelope
during its cooling
\citep{BlSorCOshells_arXiv}.

The main parameters of the best models for  SN~2010gx and PTF09cnd
are shown in the Table~\ref{table:models}.
The models starting with ``N'' (i.e., narrow) are able to reproduce the light curve of SN~2010gx,
those starting with ``B'' (broad) are constructed for PTF09cnd.
Models N0 and B0 provide the best fits,
all other models also have light curves close to the observations.
We use them to demonstrate some dependences on model parameters.
CO5, CO7, and CO9 in the column ``Composition'' mean
that the model contains roughly 50\%, 70\%, and 90\% of carbon and 50\%, 30\%, and 10\% of oxygen,
respectively.
The only model with non-uniform composition is B6.
We used it to check how much the mixing of helium with carbon and oxygen could affect SLSN light curves.
The inner 4\MS\ of the ejecta in this model is the CO mixture
with the proportion $M_{\rm C}:M_{\rm O}=9:1$, as in the model B0.
Then in the outer parts of ejecta some amount of helium appears and grows up to about 50\% in the outermost layer of ejecta.
This percentage remains fixed for the whole outer envelope.
The proportion $M_{\rm C}:M_{\rm O}$ is fixed throughout the model.

\begin{table*}
\centering                        
\caption{Model parameters (see the comments in the text).
\label{table:models}         }  
\begin{tabular}{l c c c c c c c}  
\tableline\tableline
&&&Extended&&&& \\    
Model&$M_{\rm ej}$&$p$&envelope&$M_{\rm w}$&$E_{\rm expl},$ &
$E_{\rm w,kin},$ & Composition\\    
& (\MS) && or detached & (\MS) & ($10^{51}$ergs) &($10^{51}$ergs)&\\
&&& shell &&&&\\
\tableline                        
   N0 & 0.2   &  1.8 & env & 9.7  &  2 & 0.04    &CO7 \\ 
   N1 & 0.2   &  1.8 & env & 4.9  &  2 & 0.02    &CO7 \\ 
   N2 & 0.2   &  1.5 & env & 4.8  &  2 & 0.02    &CO7 \\ 
   N3 & 0.2   &  1.8 & env & 4.9  &  2 & 0.02    &CO9 \\ 
   N4 & 0.2   &  1.8 & env & 4.9  &  2 & 0       &CO9 \\ 
   N5 & 0.2   &  1.8 & env & 4.9  &  3 & 0       &CO9 \\ 
   N6 & 0.19  &  3.5 & sh  & 9.8  &  2 & 0.1     &CO9 \\ 
   N7 & 0.19  &  3.5 & sh  & 9.8  &  2 & 0       &CO9 \\ 
   N8 & 0.19  &  3.5 & sh  & 4.7  &  2 & 0       &CO9 \\ 
\tableline
   B0 & 5     &  1.8 & env & 49   &  4 & 0       &CO9 \\ 
   B1 & 5     &  1.8 & env & 49   &  4 & 0.1     &CO9 \\ 
   B2 & 5     &  1.8 & env & 49   &  4 & 0.3     &CO9 \\ 
   B3 & 0.2   &  1.8 & env & 19   &  4 & 0       &He  \\ 
   B4 & 0.2   &  1.8 & env & 19   &  4 & 0       &CO5 \\ 
   B5 & 0.2   &  1.8 & env & 19   &  4 & 0       &CO9 \\ 
   B6 & 5     &  1.8 & env & 49   &  4 & 0       &0.5He$+$0.5CO9 \\ 

\tableline

\end{tabular}
\end{table*}

\citet{Fryer2010} applied a similar ejecta/wind structure to the DD scenario of an SN~Ia explosion, but
it is hard to imagine the formation of such an extended structure on the dynamical time-scale of
a DD event.
Nevertheless, it is interesting to consider those structures independently
of the DD scenario because they can help to explain extremely powerful SLSNe.
We discuss some feasible ways of the formation of these dense CS envelopes in Sec.~\ref{sec:summary}.

\section{METHOD OF THE LIGHT CURVE SIMULATION}
\label{sec:simul}

We  calculate the synthetic light curves
using our multi-group radiation hydrodynamic code {\sc stella} in its
standard setup \citep{BliEas1998,BliSor2004,BaklEa05,BliRop2006}.
The code simulates spherically symmetric hydrodynamic flows coupled
with multi-group
radiative transfer.
Standard runs use 300 radial Lagrangean mesh zones.
When compared to 500 mesh zones, the results
do not change significantly.
The opacity routine takes into account electron scattering, free-free,
and bound-free
processes.
Contribution of spectral lines (i.e. bound-bound processes) is treated
in approximation
of ``expansion'' opacity.
Various approaches to the expansion opacity are described in detail by
\cite{Castor2004}.
We use the method suggested by
\citet{FriendCastor1983,EastmanPinto1993}, see also
\citet{Bli1996}.

All runs employed 100 frequency groups in the transport solver and a relatively short
spectral line list ($\sim~1.5~\cdot~10^5$~lines) in the opacity routine.

 {\sc Stella} in its standard version calculates variable Eddington factors in a static approximation
once per every 50 hydrodynamic steps.
We checked
some runs with the results of the {\sc rada} code,
which computes full time-dependent
radiative transfer for intensity and uses more reliable Eddington factors
\citep{Tolstov2003,Tolstov2005,Tolstov2010}.
The comparison with {\sc stella} runs shows that the difference is not large
(see the discussion below).

The explosions have been simulated as a ``thermal bomb'' with variable
energy $E_{\rm expl}$ (see Table~\ref{table:models}).
The burst duration is 10~seconds in the innermost layers of ejecta
with $\Delta M = 0.06 M_\odot$.

\section{HYDRODYNAMICAL EVOLUTION FOR DIFFERENT MODELS}
\label{sec:hydroprofiles}

\begin{figure*}
\centering
\includegraphics[width=0.45\linewidth]{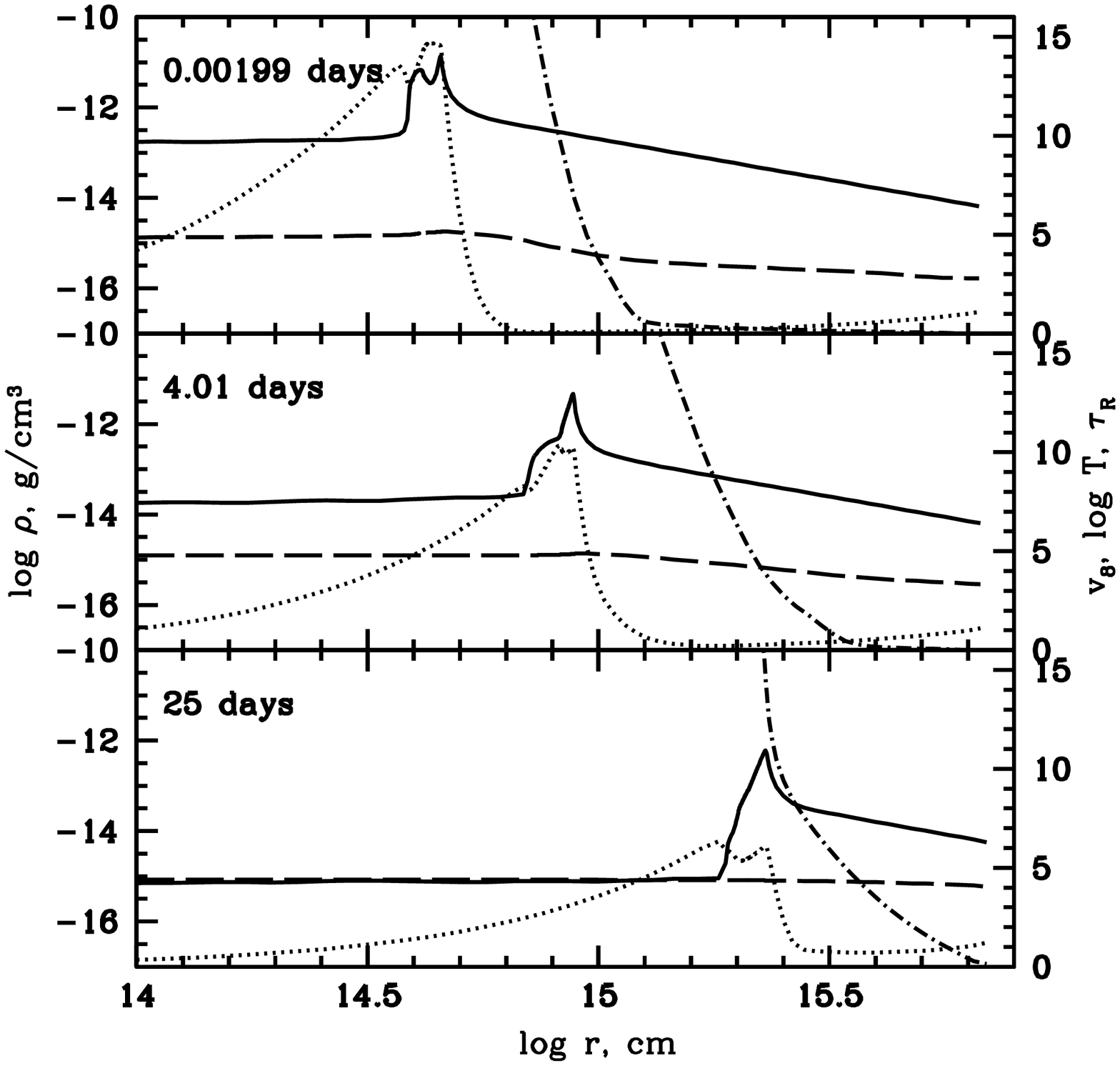}\hfill
\includegraphics[width=0.45\linewidth]{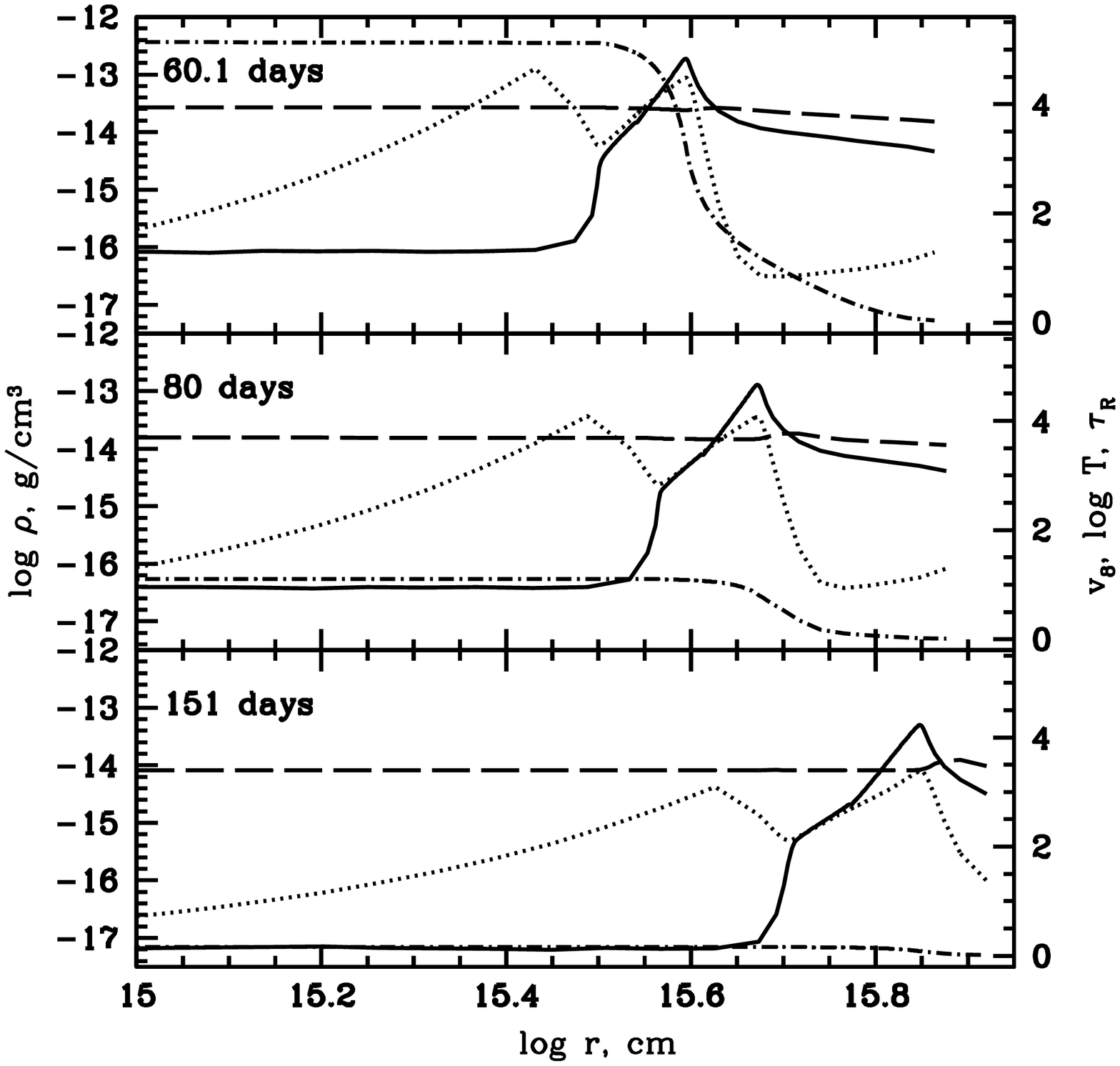}
\caption{\noindent
Evolution of radial profiles of the density ({\it solid lines}),
velocity (in $10^8$~cm~s$^{-1}$, {\it dots}), matter temperature ({\it dashes}),
and Rosseland optical depth ({\it dash-dots})
for  model N0.
The scale for the density is on the left Y axis,
for all other quantities,  on the right Y axis.
Left panel: Evolution of the hydrodynamical structure before maximum: very soon after the explosion and at days 4 and 25.
Right panel: The same parameters, but after maximum: at days 60, 80, and 151.
Note that different scales for the axes are used in the left and right panels.
        }
\label{fig:hydroN0}
\end{figure*}

\begin{figure*}
\centering
\includegraphics[width=0.45\linewidth]{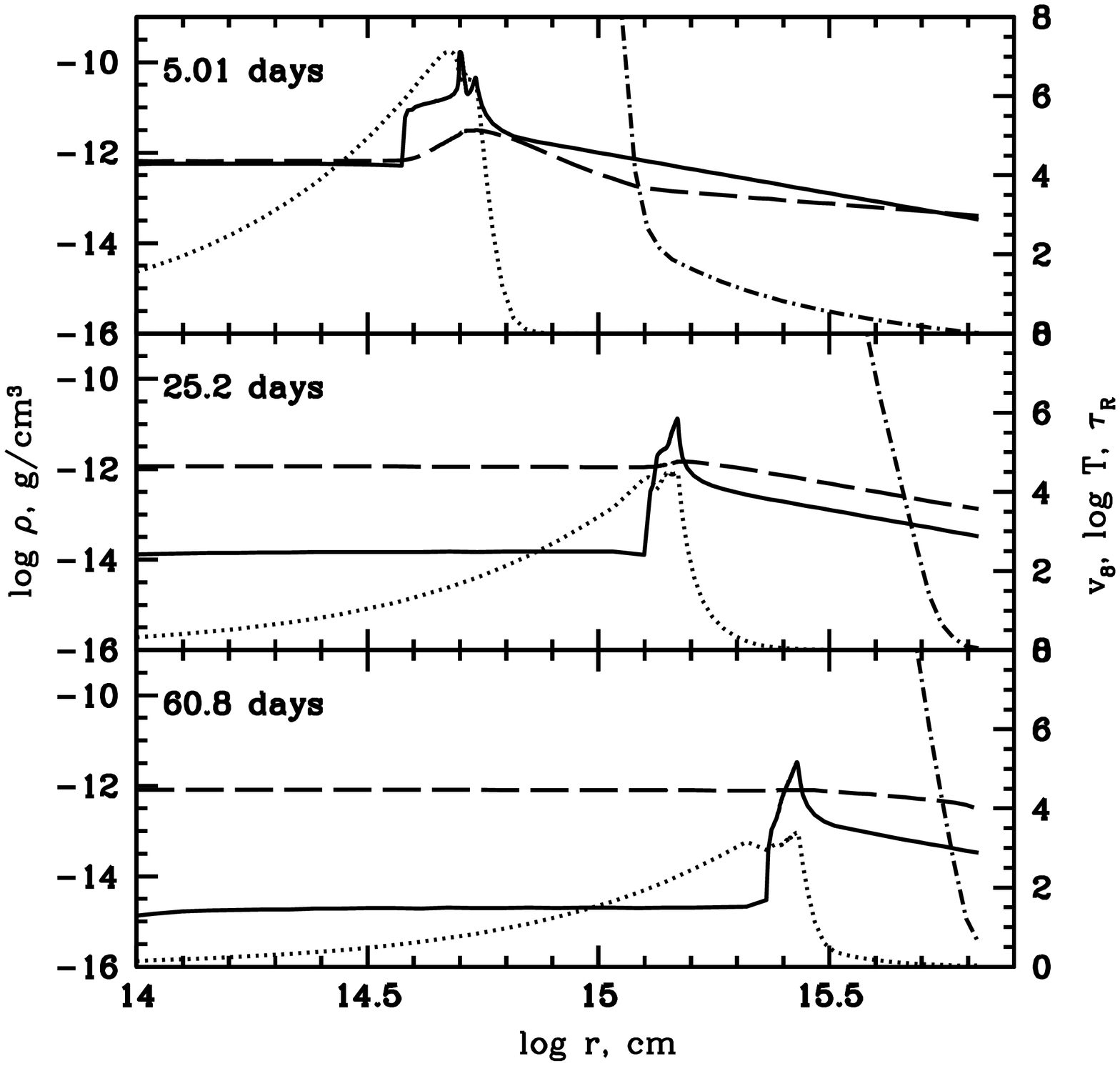}\hfill
\includegraphics[width=0.45\linewidth]{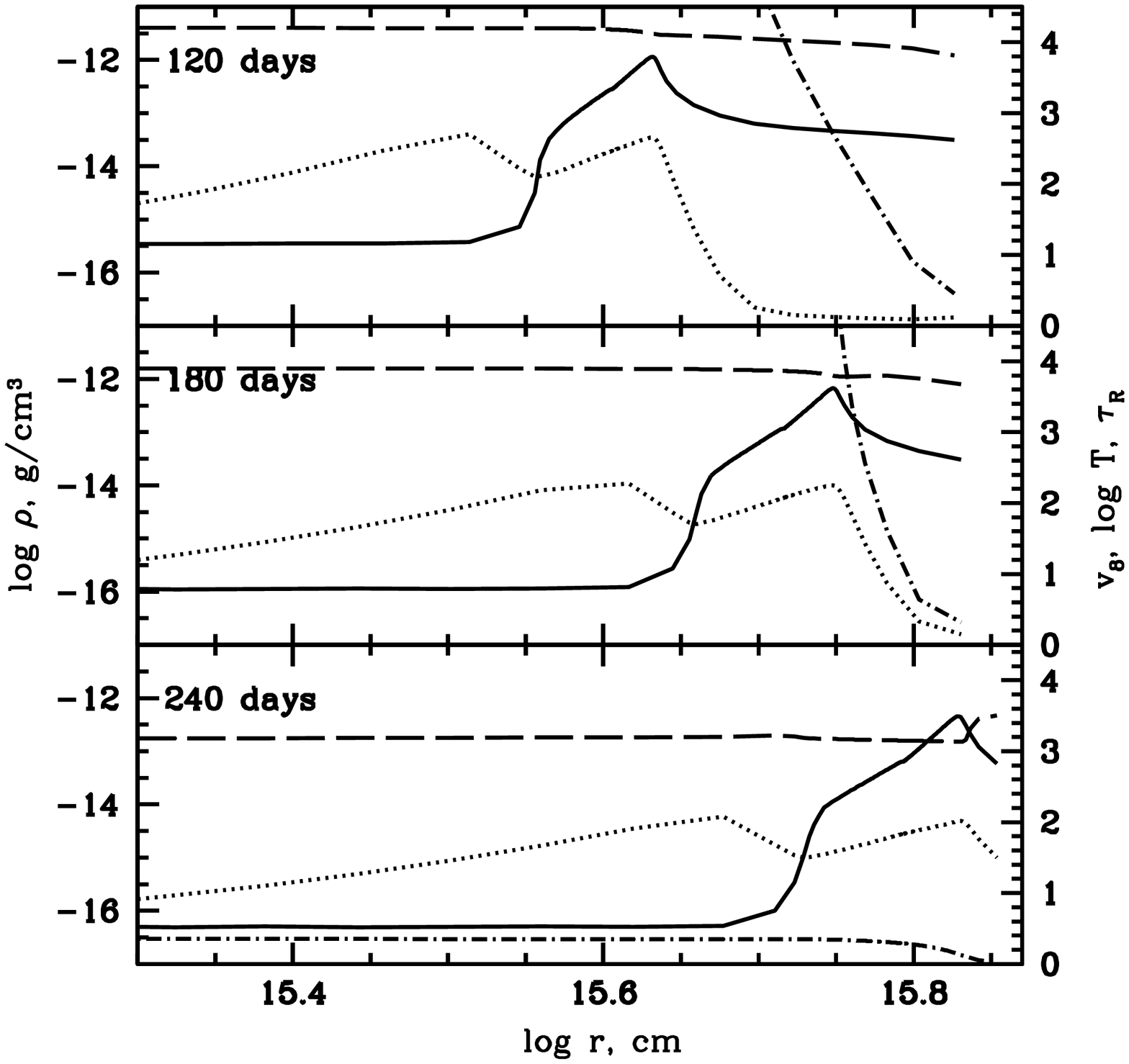}
\caption{\noindent
  Evolution of radial profiles of the density ({\it solid lines}),
velocity (in $10^8$~cm~s$^{-1}$, {\it dots}), matter temperature ({\it dashes}),
and Rosseland optical depth ({\it dash-dots}) similar to Figure~\ref{fig:hydroN0},
but for the model B0.
The ages are shown in the upper left corner  of each plot.
        }
\label{fig:hydroB0}
\end{figure*}

Figures~\ref{fig:hydroN0} and \ref{fig:hydroB0} show how the profiles of density, velocity, temperature,
and Rosseland mean optical depth evolve over time for models N0 and B0.
The left panels correspond to the evolution before maximum of the light curve
(which happens on day 22 after the explosion for N0 and on day 66 for B0 as we will show later),
the right panels show the evolution after maximum.

At the very beginning,
when the shock wave structure starts to form due to a collision between the ejecta and the CSM,
the envelope is cool and transparent (upper left plots on the Figures~\ref{fig:hydroN0} and~\ref{fig:hydroB0}).
Then the emission from the shock front heats the gas in the envelope, thus making it opaque,
and the photosphere moves to the outermost layers rather quickly.
When the photospheric radius reaches its maximum,
one can observe maximal emission from the supernova.

The speed of the growth of the photospheric radius depends on the mass of the envelope,
since more photons must be emitted from the shock to heat  larger mass envelopes.
This is why the photosphere moves to the outermost layers in model N0 faster than in model B0
(the plots on the left-hand side of Figures~\ref{fig:hydroN0} and \ref{fig:hydroB0}).

\begin{figure}

\centering
\includegraphics[width=0.45\linewidth]{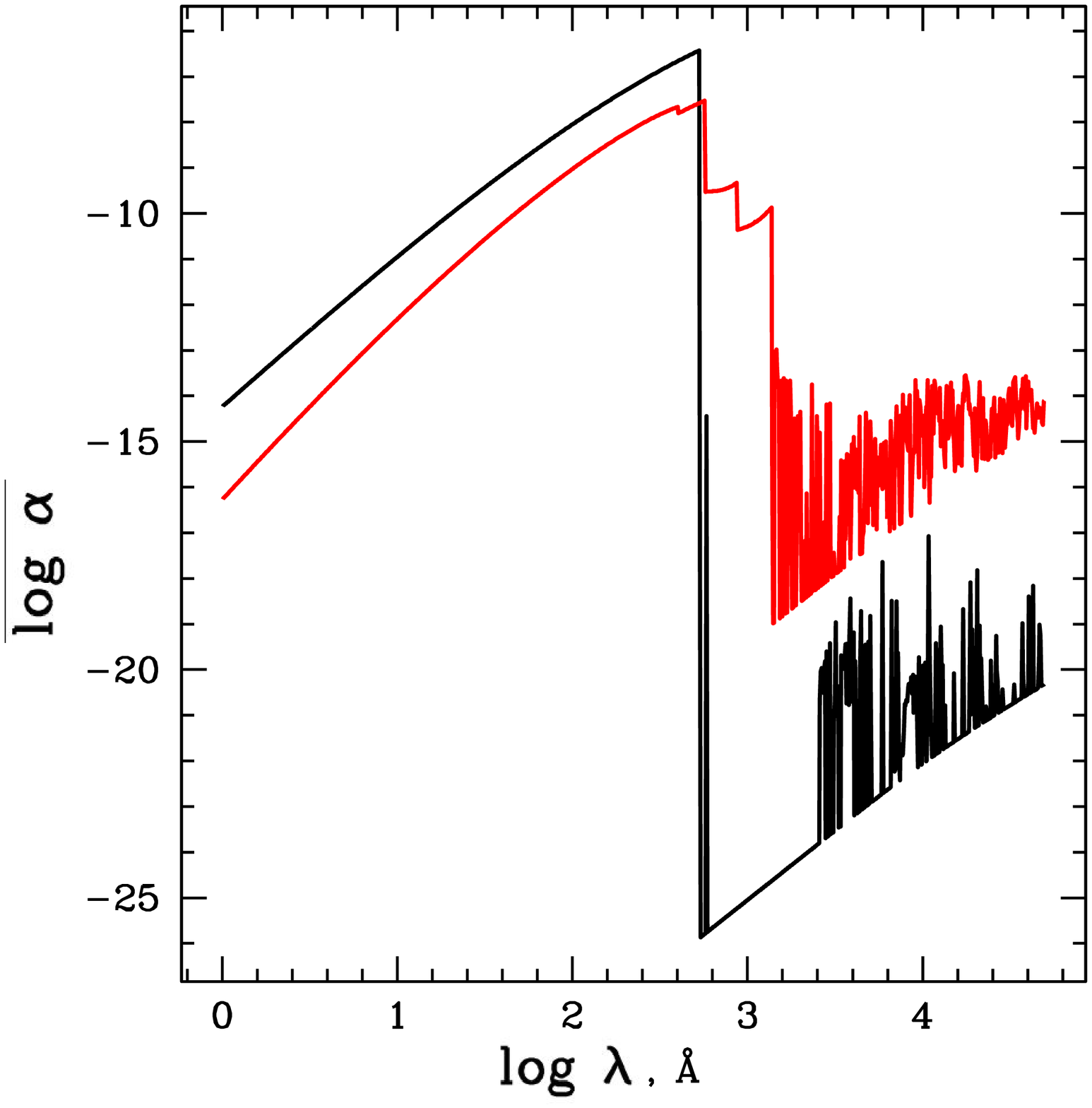}
\includegraphics[width=0.45\linewidth]{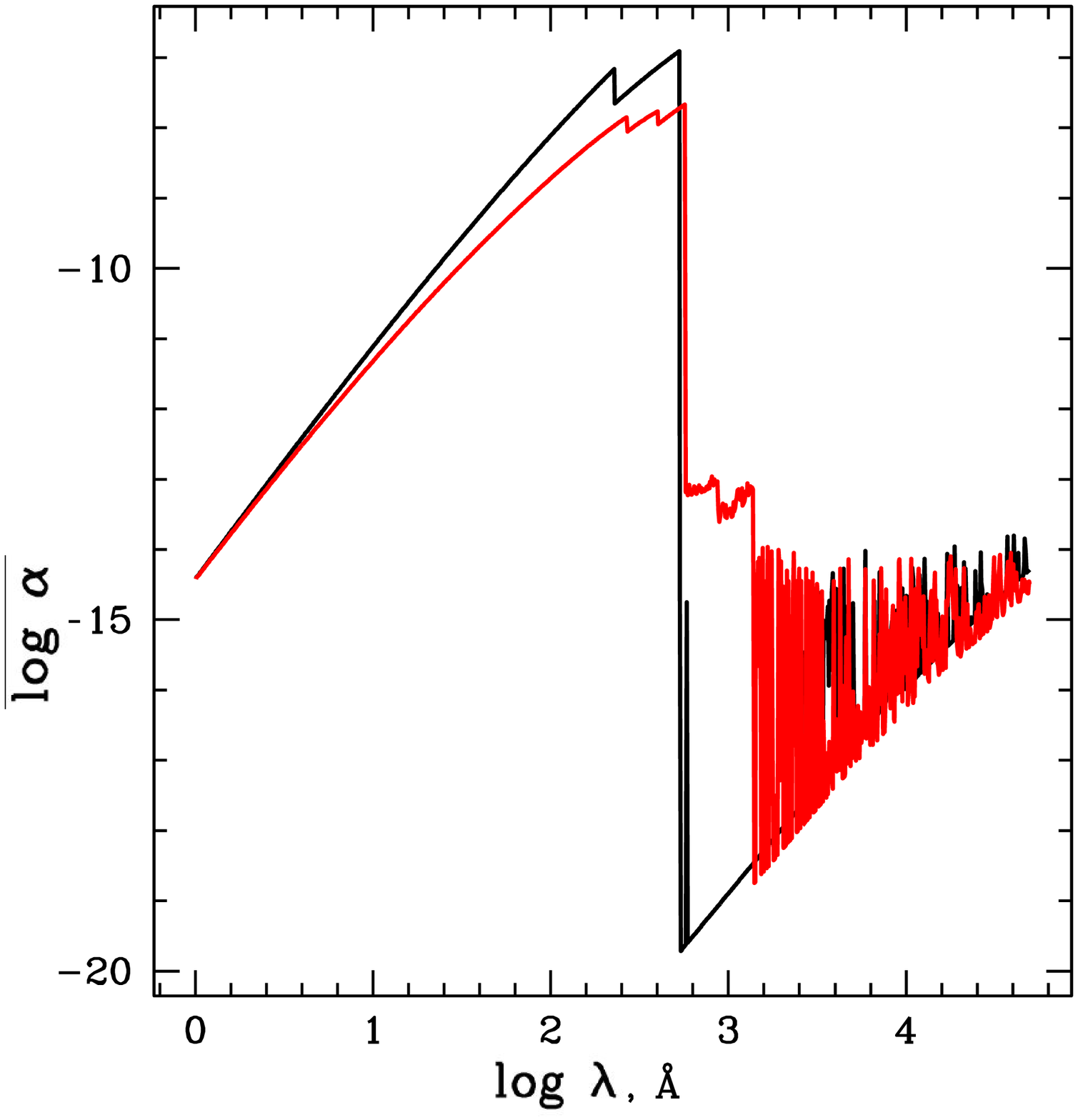}
\caption{\noindent Expansion opacity for helium ({\it black lines})
and mixture of 90\% carbon and 10\% oxygen ({\it red lines})
for $T=7000$~K ({\it left panel}) and $T=11000$~K ({\it right panel}).
        }
\label{fig:opaHe_CO9}
\end{figure}

Another parameter that impacts the initial growth of the photospheric radius
is the chemical composition of the envelope.
In Figure~\ref{fig:opaHe_CO9} we compare the opacities of pure helium
and mixture of carbon (90\%) and oxygen (10\%) for the same density $3\cdot 10^{-13}$~g/cm$^3$.
The left plot shows opacities at the temperature  7~000~K.
At this temperature, helium still remains transparent, so the photosphere is deep inside the envelope.
The opacity for the CO mixture under the same conditions is six orders of magnitude higher.
The optical depth at 7~000~K is large, so one can see only the outermost layers of the CO envelope.
At 11~000~K (right plot) the helium opacity rises up and becomes similar to the CO one.
In this case, the photospheric radii for both compositions must also be similar.
Thus, we predict that  the light curve rises faster for a CO envelope than for a He one
because a lower temperature is needed to reach high opacity in a CO mixture.
This light curve behavior can help set the composition for some observed SLSNe.

The plots on the right-hand side of Figures~\ref{fig:hydroN0} and \ref{fig:hydroB0} show  the stages
when the photosphere slowly moves back to the center,
and the envelope and the ejecta finally become fully transparent.
At the beginning of this post-maximum stage, all gas in the envelope is already heated by the photons,
which came from the shock region and diffused through the envelope to the outer edge,
and the whole system (ejecta and envelope) becomes almost isothermal.
The shock becomes weaker with time and emits fewer photons that can heat up the envelope,
so the temperature of the still unshocked envelope falls down.

The shocked material is gathered into a thin, dense layer,
which finally contains almost all the mass in the system.
The formation of this layer leads to numerical difficulties,
which significantly limit the time step of the calculation.
Due to the fixed Lagrangean grid,  most of the spacial bins in the layer have very similar radii,
but their densities and temperatures are a bit different.
Thus, gas does not cool down identically in each bin,
and counter motions develop in the dense layer
when the pressure of neighboring spacial bins flattens out.
This can lead to the overlapping of some bins
if the time step of the calculation is too large.
Short time steps are required to avoid this overlapping.
The use of an adaptive spacial grid might help,
but {\sc stella} has no such  possibility.
Another problem can also take place due to the thin layer formation:
a thin, dense shell with a very large radius would most probably be unstable and can fragment into smaller clumps.
Then the problem would become essentially multi-dimensional.

On the velocity profiles, the multi-reflection structure forms from the very beginning.
It evolves very quickly to the standard two-shock (forward and reverse) picture.
This does not depend on the initial velocity profile in the envelope.
The interaction of the ejecta with an expanding (model N0) or static (model B0) envelope leads to similar
final velocity structures.
It looks like a self-similar behavior analogous to the solution by \citet{Nad1981,Nad1985,Chevalier1982},
but with radiation.


\section{LIGHT CURVES AND SPECTRA}
\label{sec:LCsp}

\subsection{General properties of the interacting SLSN light curves}
\label{sec:res-general}

We have chosen two type I SLSNe to model -- SN~2010gx ($\equiv$~PTF10cwr) and PTF09cnd.
Both were described among the PTF set of SLSNe in \citet{QuimbyNatur2011} \citep[see also][]{Pastorello2010}.
SN~2010gx and PTF09cnd have some of the narrowest and the broadest light curves, respectively, among all SLSN-I.
Our goal is to model both of them by the interaction of the ejecta with the surrounding envelope
in order to understand if most SLSN-I can be explained by the same mechanism
with different parameters of either explosion, or CSM.

We are interested in studying a pure effect of interaction with CSM for the SLSN light curve,
thus we do not add \nifsx or any other additional energy source inside our models.
All of the emission comes from the transformation of ordered particle motion into a chaotic state
when gas passes through a shock wave.
When the shock reaches the outer edge of the extended envelope and no material remains in front of the shock anymore,
no source of energy remains.
The gas looses its thermal energy through radiation and cools down very quickly.
It corresponds to a sharp drop in flux in all spectral bands.
This drop is a typical feature of the light curves for the interacting models,
though it must not  necessarily be seen on the observed SLSNe
since it happens a few months after maximum (if the envelope is extended enough).
The supernova may be unobservable or a second energy source may become dominant by this phase.

One clearly needs a very large radius envelope to produce an extremely bright and long-lasting event
for a model without a huge explosion energy.
One also needs high densities for strong production of
light by the shock.
However, when the density is too high,
the mass of the envelope
and the optical depth of the shell become too large.
This would make the supernova appear red
and would not match with observations of SLSNe-I, which tend to be blue \citep[e.g.][]{QuimbyNatur2011}.
Thus, an enhanced envelope mass must be accompanied by an enhanced explosion energy,
which will lead to the formation of a stronger and hotter shock.

We have run tens of different models, varying many parameters.
We will provide a detailed discussion on the wide set of models
and analyze the influence of the model parameters on the bolometric and broad band light curves
in a future paper (E.Sorokina et al. 2016, in preparation).
Here we show only a small part of our results,
which relates to the light curves of the SLSNe similar to SN~2010gx and PTF09cnd.

\subsection{SN~2010gx}
\label{sec:res10gx}

\begin{figure}

\centering
\includegraphics[width=0.7\linewidth]{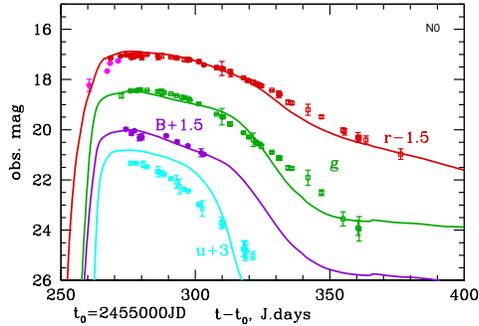}
\caption{\noindent Synthetic light curves for the model N0, one of the best for SN~2010gx, in $r$, $g$, $B$, and $u$ filters compared with Pan-STARRS and PTF observations.
Pan-STARRS points are designated with open squares ($u$, $g$, and $R$ bands)
and PTF points with filled circles ($B$ and $r$ bands).
Four pink points in the beginning of the $r$ band shows PTF observations in the Mould $R$-band
which is similar to the SDSS $r$ band.
        }
\label{pic:best10gx}
\end{figure}

\begin{figure*}
\centering
\includegraphics[width=0.45\linewidth]{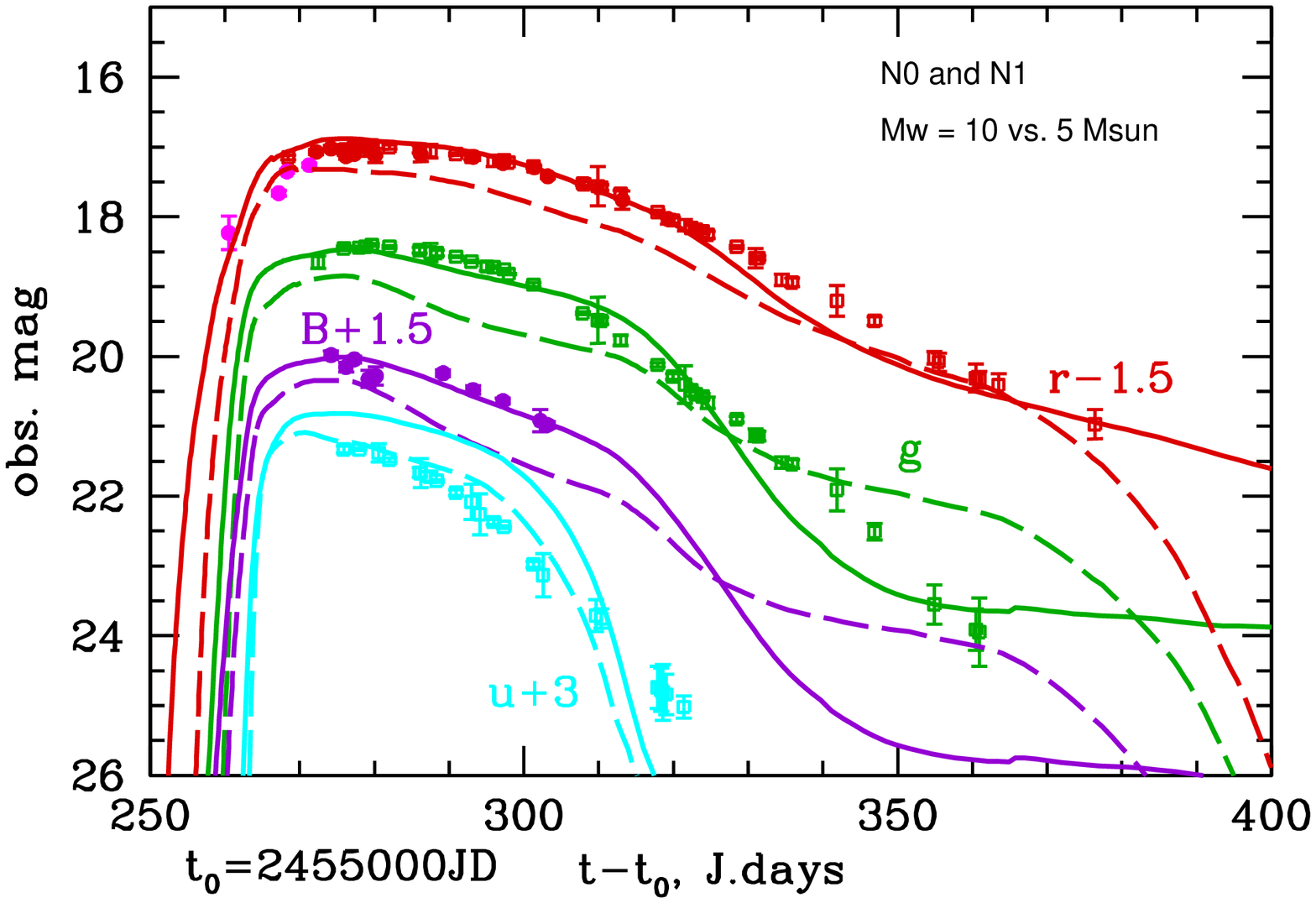} \hfil
  \includegraphics[width=0.45\linewidth]{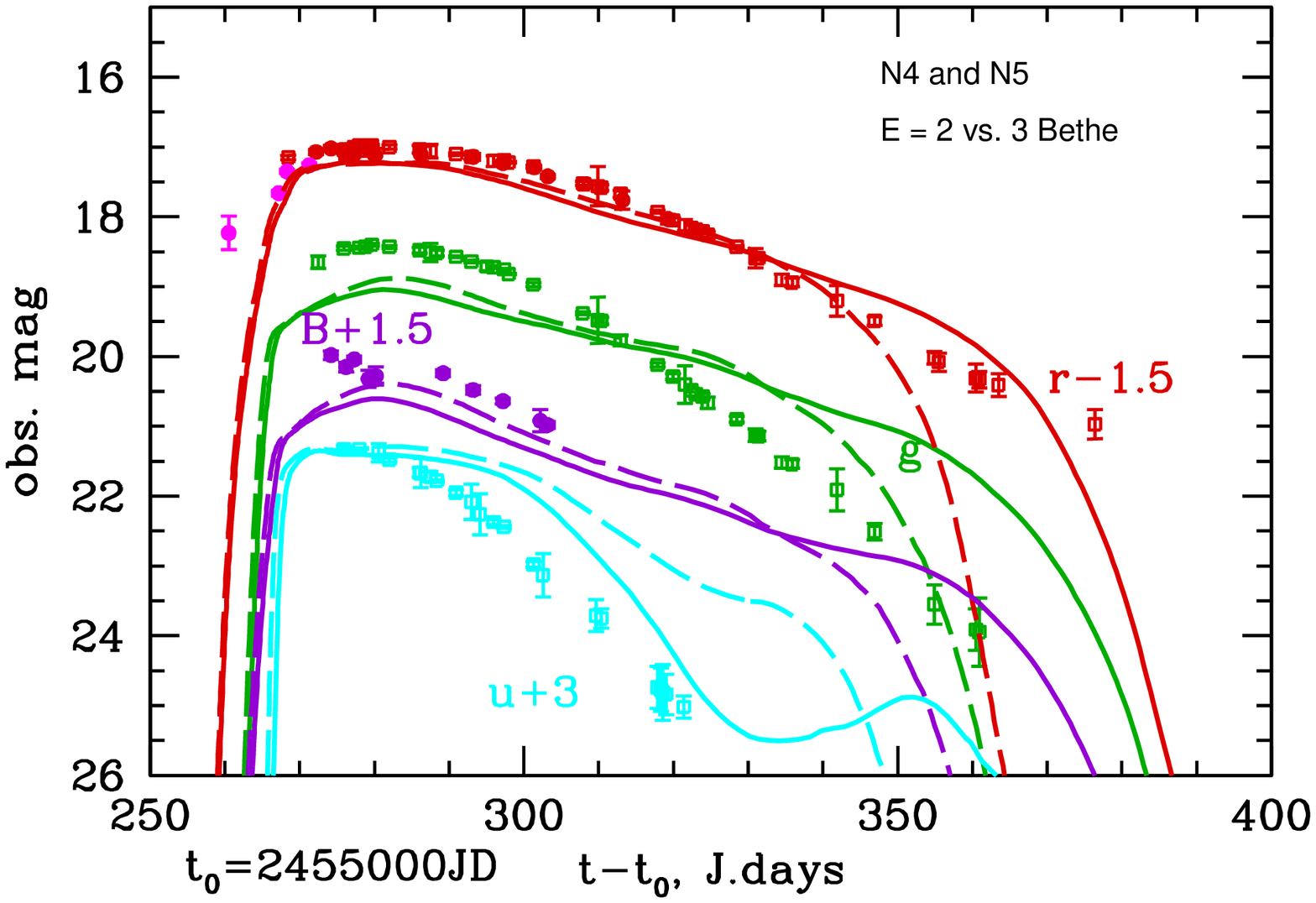}\\
\includegraphics[width=0.45\linewidth]{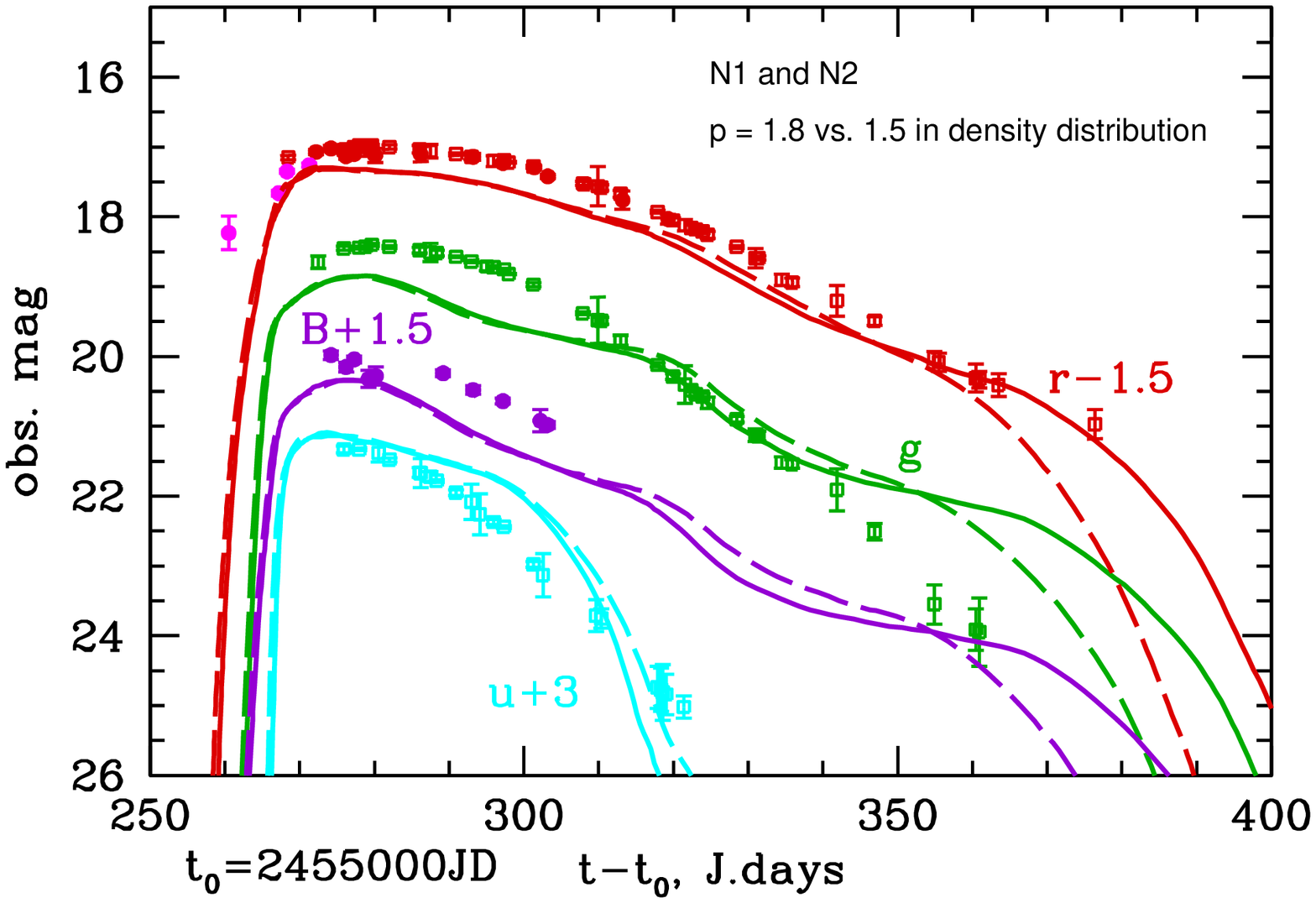} \hfil
  \includegraphics[width=0.45\linewidth]{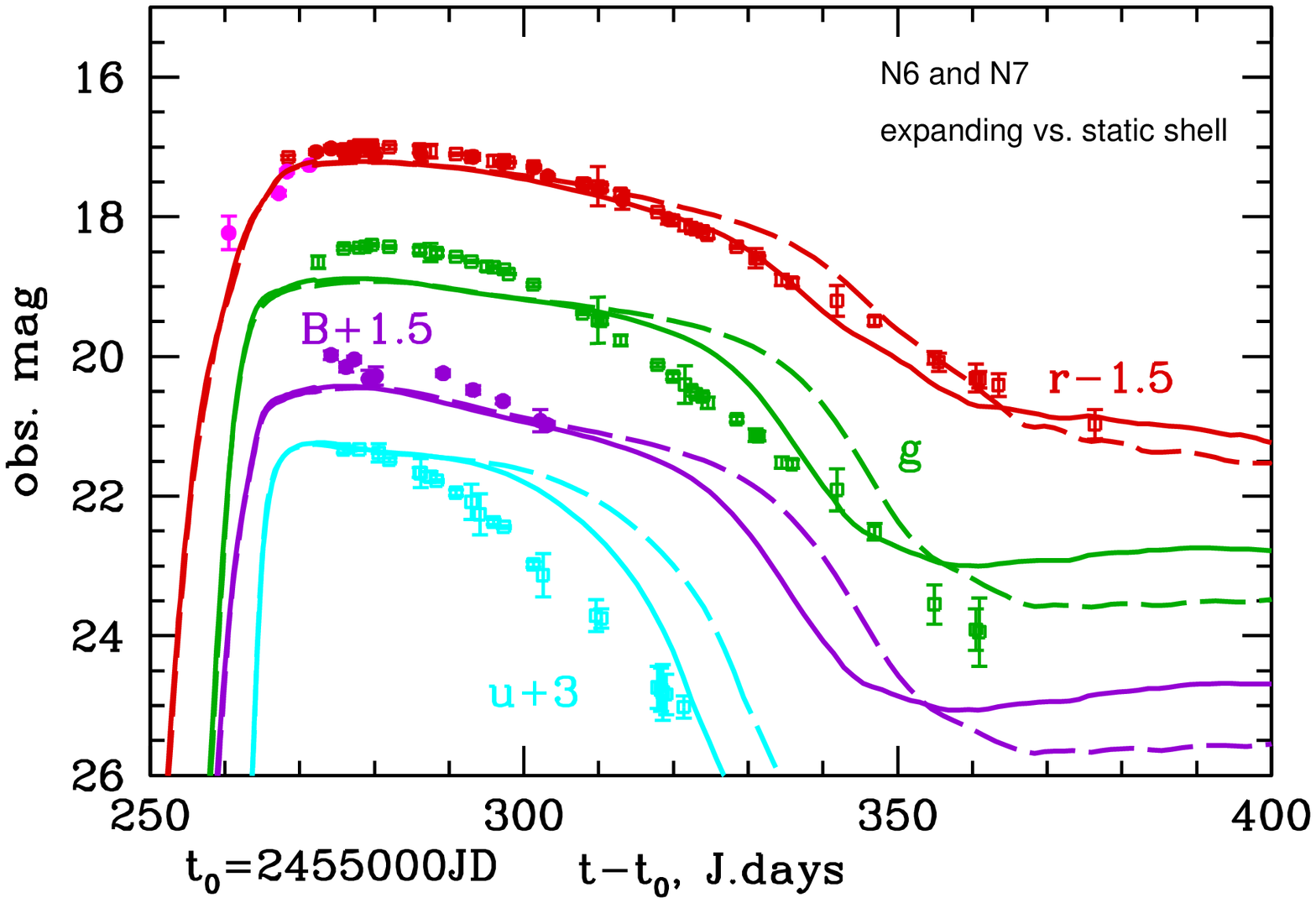} \\
\includegraphics[width=0.45\linewidth]{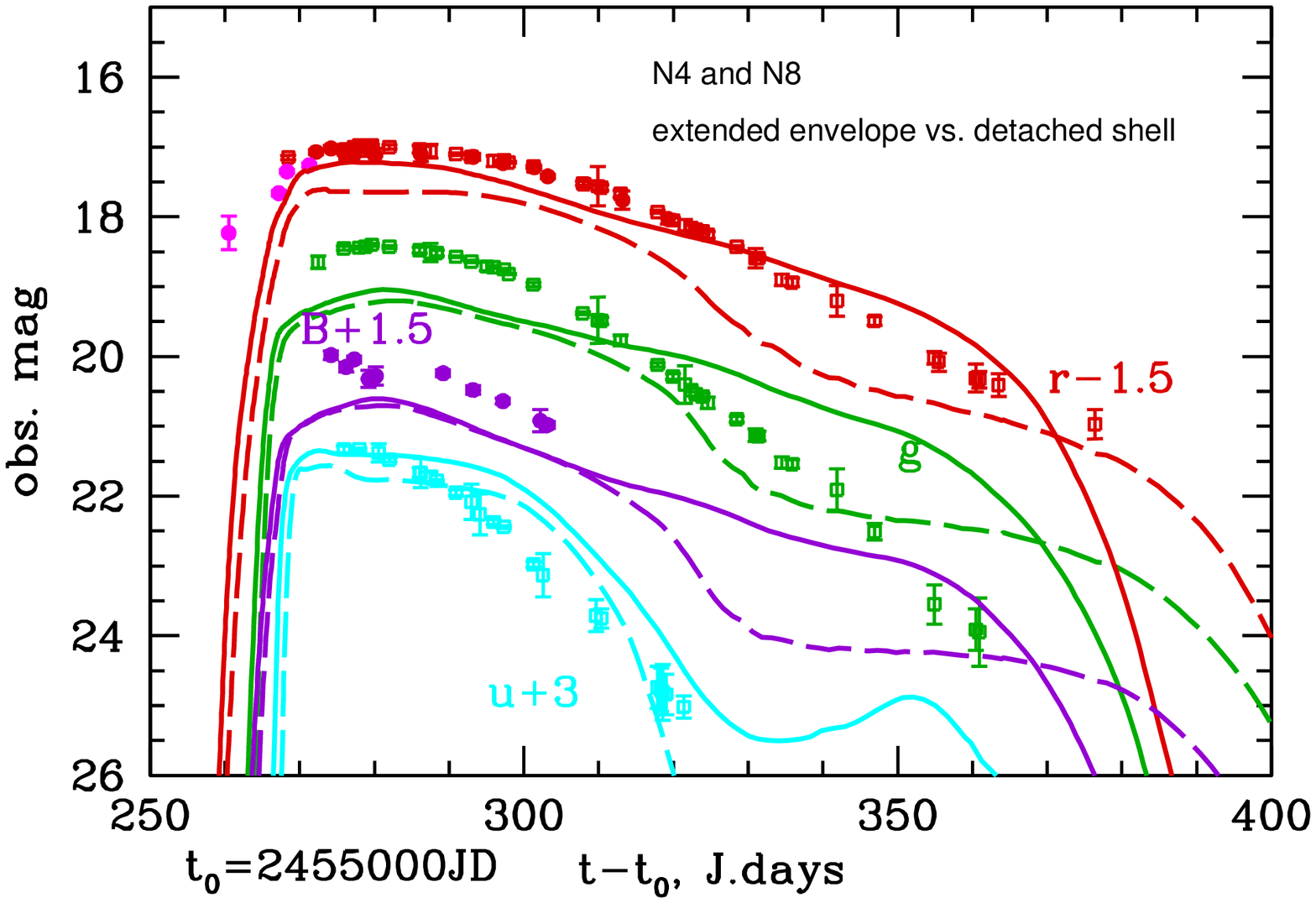} \hfil
  \includegraphics[width=0.45\linewidth]{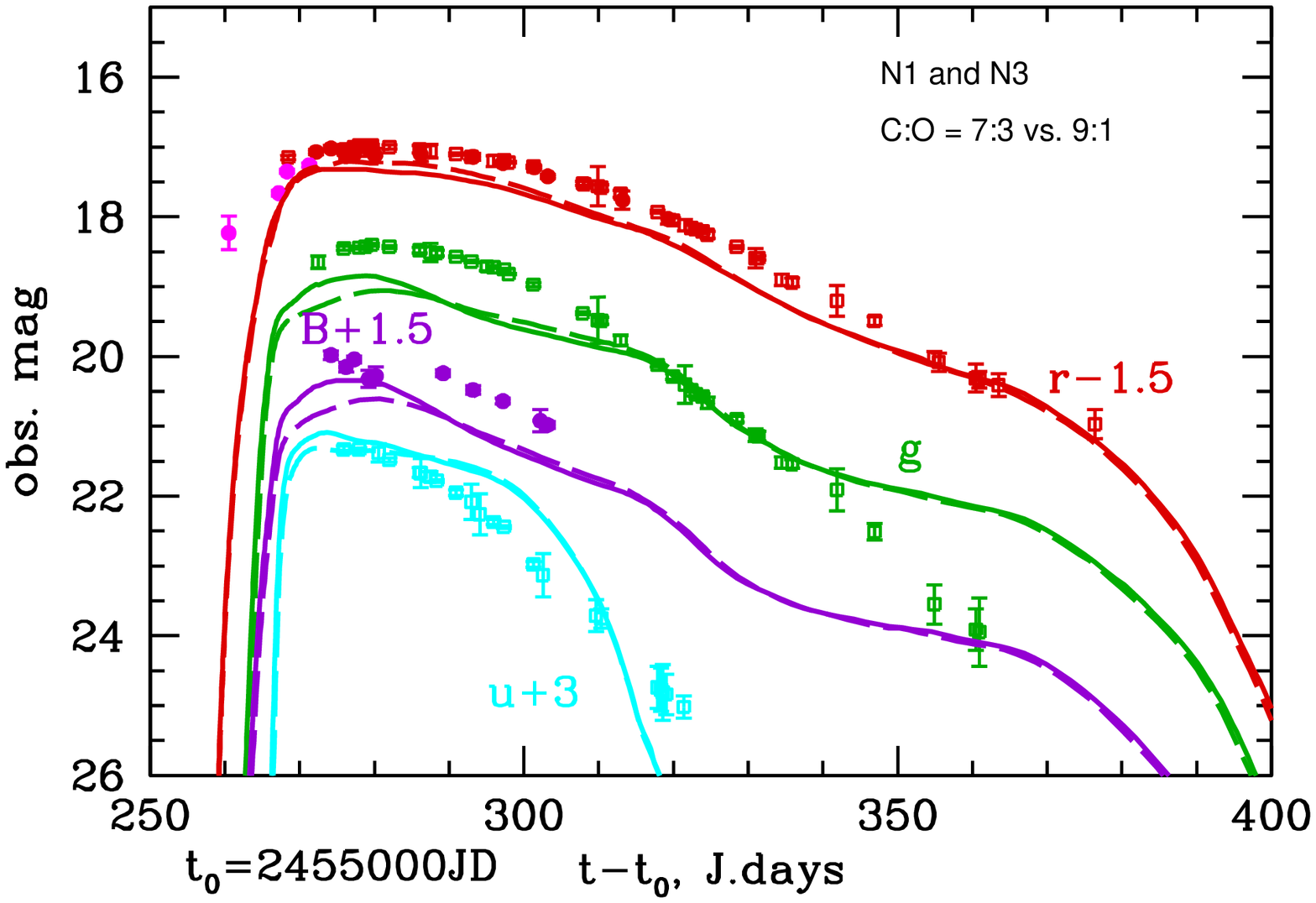}
\caption{\noindent Dependences of the light curves in $uBgr$ filters on various model parameters
are shown by comparing a relevant pair of models in Table~\ref{table:models}.
Squares and circles designate observational points for SN~2010gx in the same way as in Figure~\ref{pic:best10gx}.
The {\it upper left} plot compares the light curve for the models N0 and N1 with different envelope mass,
  10~\MS ({\it solid}) and 5~\MS ({\it dashed});
{\it upper right:} models N4 and N5 with different explosion energy,
  $2\cdot 10^{51}$~ergs ({\it solid}) and $3\cdot 10^{51}$~ergs ({\it dashed});
{\it middle left:} models N1 and N2 with different power for density distribution in the envelope,
  $\rho\sim r^{-1.8}$ ({\it solid}) and $\rho\sim r^{-1.5}$ ({\it dashed});
{\it middle right:} model N6 with an expanding detached shell with total kinetic energy $E_w=10^{50}$~ergs
  and the highest velocity about 1760~km~s$^{-1}$ ({\it solid})
  and model N7 with static detached shell ({\it dashed});
{\it lower left:} models with different density distribution in the envelope,
  N4 with power law $\rho\sim r^{-1.8}$ ({\it solid})
  and N8 with the gas concentrated into the detached shell ({\it dashed}) having the same mass as the power-law model;
{\it lower right:} models N1 and N3 with different abundance ratio C:O, 7:3 ({\it solid}) and 9:1~({\it dashed}).
        }
\label{pic:10gx_parameters}
\end{figure*}

We have already presented some models for SN~2010gx in \cite{BlSorCOshells_arXiv}.
Since that time, several parts of the code {\sc stella} have been updated,
and the results have slightly changed.
In the current work, we present models that better reproduce the light curves of SN~2010gx.

So far, no numerical calculations except for ours were presented for the broad band light curves of SLSNe-I
in the ejecta-CSM interaction scenario.
{\sc stella} allows us to produce light curves
in different broad band filter systems.
Here we present $ugri$  light curves in the AB system \citep{OkeGunnABmags}
as well as {\it B}-band light curves normalized by Vega magnitudes
in order to compare the models with Pan-STARRS 1 \citep{Pastorello2010}
and Palomar Transient Factory \citep{QuimbyNatur2011} data.
We adopt standard cosmology ($H_0=71$~km~s$^{-1}$~Mpc$^{-1}$, $\Omega_m=0.27$, and $\Omega_\Lambda=0.73$) and $z=0.23$
for SN~2010gx to transform the modeled absolute magnitudes to observational ones.
This corresponds to the distance modulus 40.28.

\subsubsection{The best model}
\label{subsec:best10gx}

Model N0 seems to provide the best match to SN~2010gx (Figure~\ref{pic:best10gx}).
The maximum emission and the slopes of the light curves after the maximum are in excellent agreement,
though the light curve in the {\it u} band is a bit too bright.
However, the opacity is large in the u-band and it is possible
that the observed flux has been absorbed in this band along the line of sight.
We thus do not take the brighter model predictions as necessarily inconsistent with the data.
If extinction in {\it u} along the line of sight is not sufficient to put the model into agreement with observations,
then we have to change the composition for those elements in the model that can influence mostly the {\it u} band.

Deviation from the observed light curves in {\it g} and {\it r} bands 2 months after the maximum
may be caused by a more complicated structure of the envelope than we  use in the models.
We do not attempt to fit every detail of the light curves, since
there are many numerical uncertainties.
For example, switching to a more realistic velocity gradient in the expansion opacity calculations
could affect fluxes in the bands with a large number of strong lines.
Moving to a 3D calculation  would introduce a viewing angle dependence in the observed flux.
We will discuss some uncertainties in the last section of this paper.

\vfill

\subsubsection{Dependence on model parameters}

Study of the influence of model parameters on the details of the SLSN light curves
is very important for better understanding of the nature of these objects
and useful in the modeling of any future SLSNe.

We examined many models while simulating the light curves for SN~2010gx.
In the current section, for each model parameter we choose a pair of models that are
identical in all but one that model parameter.
The chosen models are independent and not the variations of N0.

Parameters of the chosen models are presented in Table~\ref{table:models}.
Their light curves  are shown in Figure~\ref{pic:10gx_parameters}.
Our aim of presenting this Figure is to demonstrate parametric dependences of the light curves of SLSNe,
so the light curves of SN~2010gx are plotted only for the purpose of calibration.

Let us discuss the specific details of the light curves and the parameters that are important for those details.

Variations of the envelope mass and the explosion energy influence both the maximum and the tail parts
of the light curve (two upper plots of Figure~\ref{pic:10gx_parameters}).

{\it Mass of the envelope.}
The mass of the envelope has been changed by the variation of the envelope density within the given radius.
The emission of the higher mass model N0 is  produced by gas with higher density, and the model is thus brighter.
The shock wave needs more time to pass through the denser envelope,
so it lives longer and the light curve tail lasts longer until the shock reaches the edge of the envelope.

{\it Explosion energy.}
The main effect of increasing the explosion energy is to shorten the light curve tail:
the shock is stronger and passes through the envelope faster.
One can also expect an increase of brightness for the more energetic model,
but the envelope is essentially diffusive, the shock is deep inside the photospheric radius,
and photons that come from the level of the photosphere are not much hotter than in the less energetic case.

{\it Expansion of the envelope.}
The second row plot on the right hand side of Figure~\ref{pic:10gx_parameters} compares the light curves
for the cases of static and expanding detached shells (models N7 and N6).
The light curves for both models coincide in the beginning, while  the models are almost identical,
but later on the emitting gas in N6 becomes less dense  due to the expansion,
so it becomes transparent and produces a sharp drop in luminosity earlier
than in static model N7.
On the other hand, the shock wave in the expanding model becomes directy visible at an earlier stage --
when it is hotter;
thus, after the early drop of radiation, the emission rises up again to a higher value than in the static case.

{\it Density structure.}
The plots in the lower two rows on the left hand side of Figure~\ref{pic:10gx_parameters}
demonstrate dependences of the light curves on the density structure within the envelope.
There is indirect evidence that the density profiles around supernovae
with strong circumstellar interaction may be shallow in comparison to the case of steady wind \citep{Prieto2007},
or have rather complicated structures due to presupernova evolution \citep{Dwar2010}.
We compare different slopes $p$ in density distribution (models N1 and N2)
as well as monotonic and shell-like distribution (models N4 and N8).
As we can see from the plots, variation in the density distribution
can help to reproduce features of the observed light curves during the fading stage.

{\it C/O ratio.}
The last (lower right) plot of Figure~\ref{pic:10gx_parameters}
compares the light curves of models with different C/O ratios.
While the previous parameters changed mostly the fading part of the light curve,
this difference, on the contrary, influences the stages near maximum light.
The presence of oxygen (in our case, increasing the O fraction by three times) makes the model noticeably bluer
near maximum light.
Oxygen has strong ultraviolet lines, which increase the emissivity  in the u-band.

We plan to investigate some SLSN-I light curve dependencies on the model parameters in more detail
in the next paper.


\subsection{PTF09cnd}
\label{sec:09cnd}

\subsubsection{Observational properties}

\begin{figure*}
\centering
\includegraphics[width=0.3\linewidth]{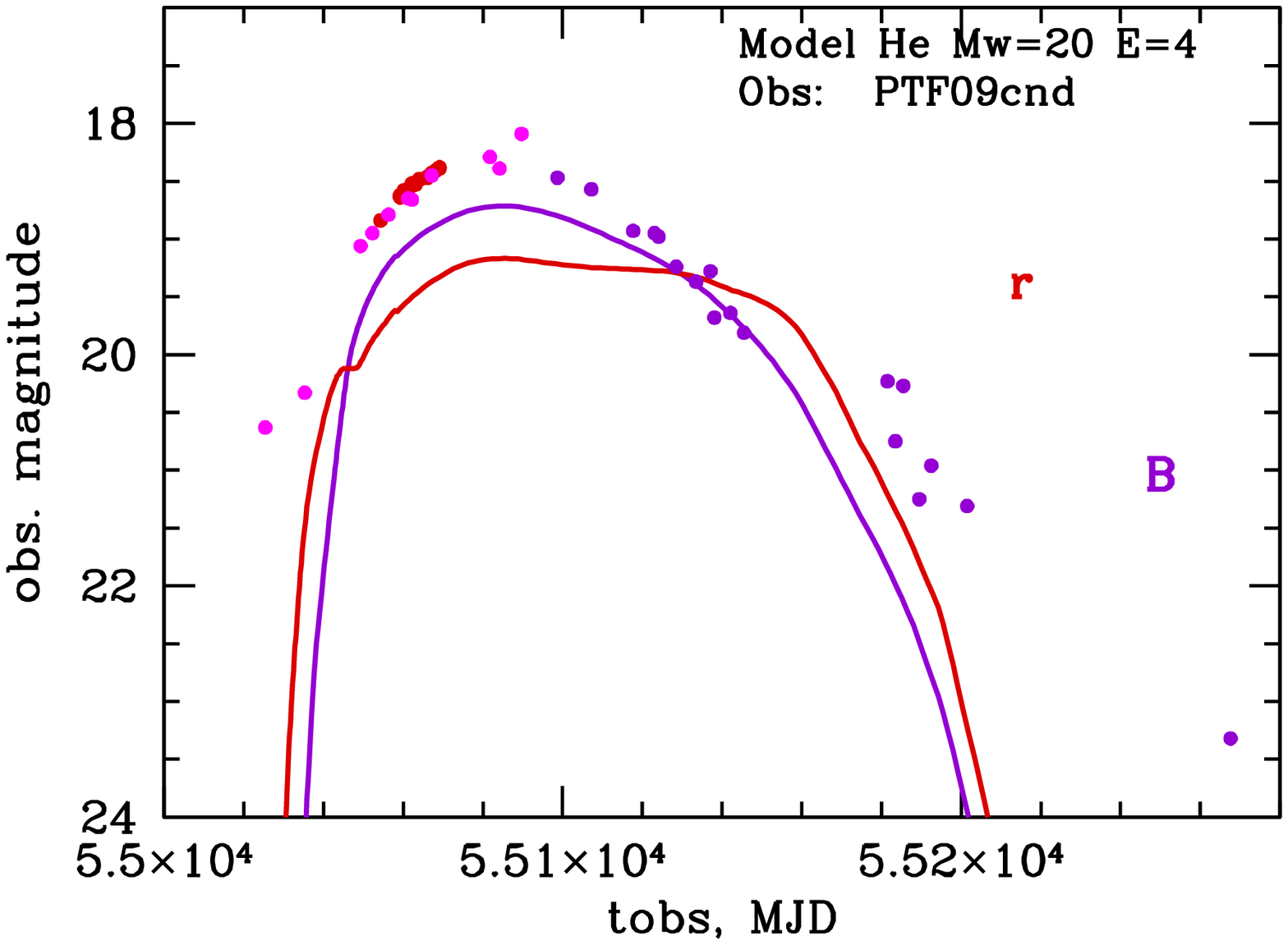} \hfil
   \includegraphics[width=0.3\linewidth]{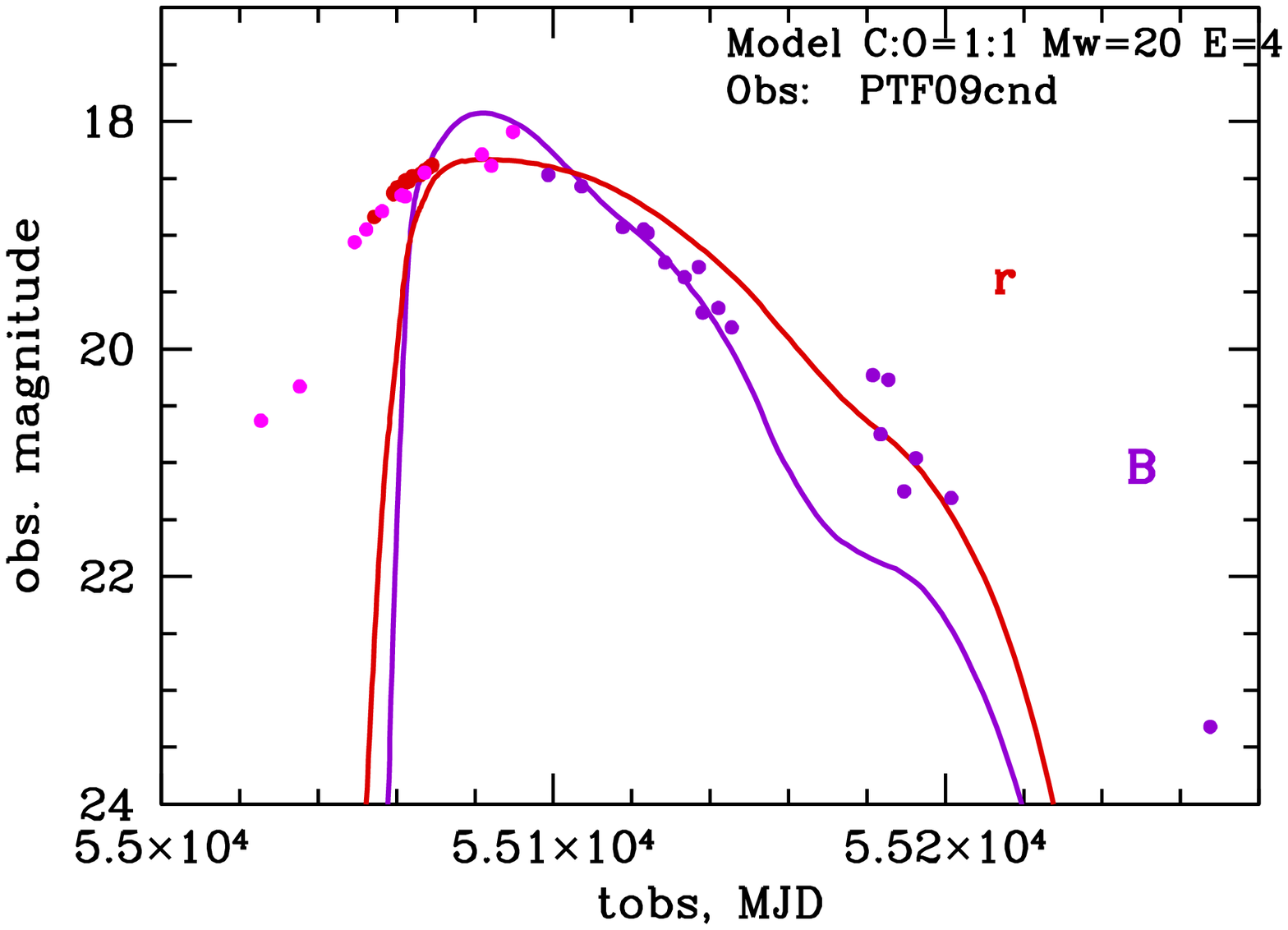} \hfil
      \includegraphics[width=0.3\linewidth]{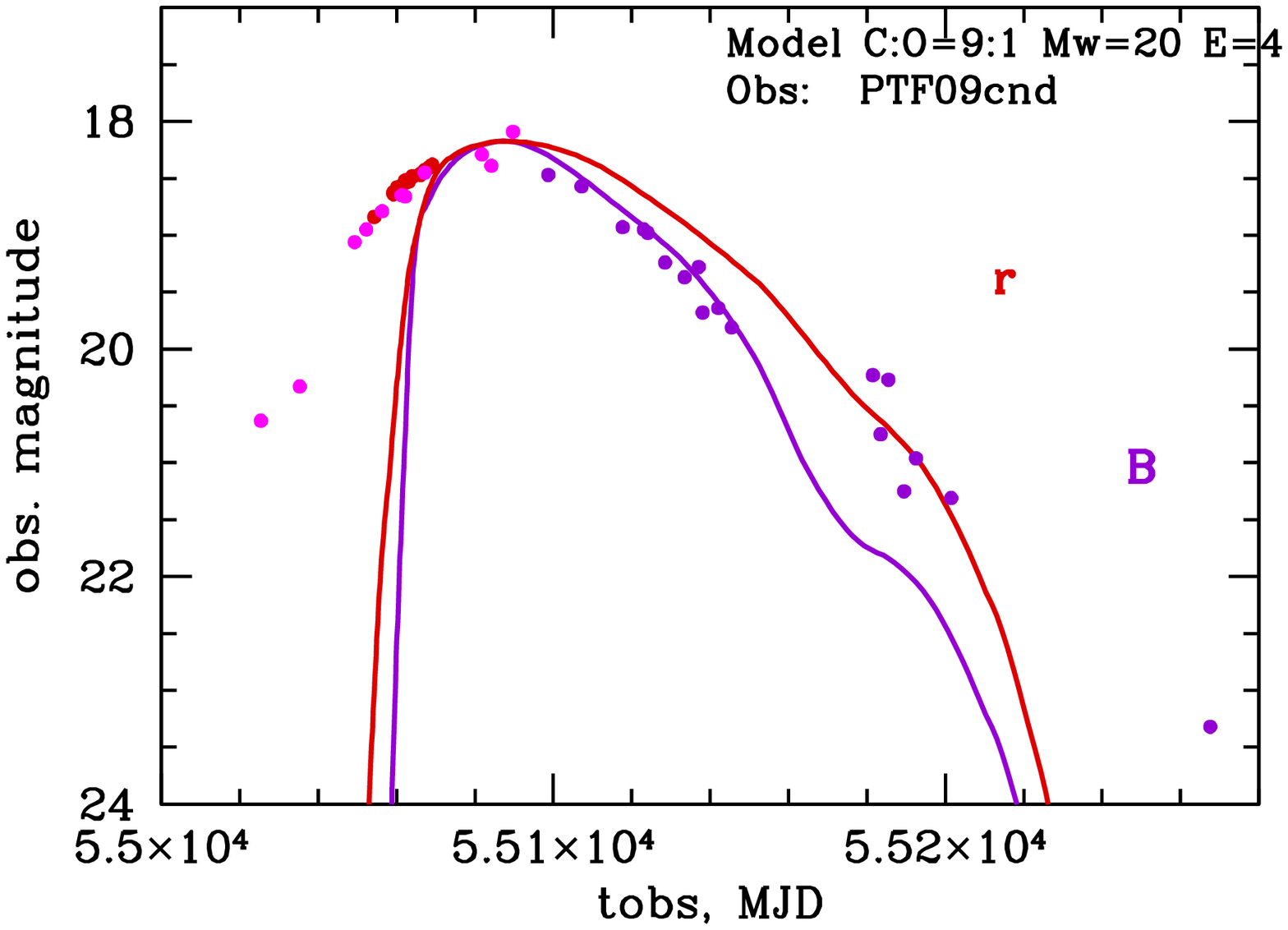}
\caption{Comparison of helium and CO (with 50\% and 90\% of carbon) models of 20~\MS (B3, B4, and B5).
       }
\label{pic:09cnd_composition}
\end{figure*}

There are several features that distinguish PTF09cnd from SN~2010gx observationally.
PTF09cnd is SLSN-I with one of the broadest light curves.
The flux in the {\it B} band drops down after maximum much more slowly than for SN~2010gx.
The maximum itself is  a little brighter than for SN~2010gx.
An additional and very important feature of PTF09cnd is that
it was caught more than 50~days before maximum,
so we have a well-documented rising part of the light curve for theoretical modeling.
We have no other photometric data for PTF09cnd except for the PTF data,
so we compare only the flux in {\it r} and {\it B} bands provided by PTF observations.
In the plots for the  observed magnitude vs. observed time, we have taken into account the redshift
of PTF09cnd ($z=0.258$).

There are two features of the PTF09cnd light curve that limit us most of all
in choosing the composition of the envelope.
The first one is the fact that maximum magnitudes
in {\it r} and {\it B} bands are very similar.
The second is its long rising time.

\subsubsection{Rising time and color at maximum in the models}

For the models with  massive envelopes,
the rising time is determined by the diffusion time of the photons from the shock wave
and by the movement of the photosphere to the outermost edge of the envelope.
We have already seen in Figures~\ref{fig:hydroN0} and \ref{fig:hydroB0} of Section~\ref{sec:hydroprofiles}
that the envelope is cool at early times.
It  is transparent for optical photons and the photosphere is deep
in the innermost part of the envelope.
However, the radiation dominated shock forms rather quickly.
Its radiation ionizes gas in the envelope, the envelope becomes opaque,
and the photosphere moves to the outer layers.
If the envelope is extended enough and the shock wave propagating through the envelope is strong enough,
the photospheric radius can grow by a few orders of magnitude,
which corresponds to the rising part of the light curve.
Time needed for this rise depends on the chemical composition of the envelope
(see Figure~\ref{fig:opaHe_CO9} and explanation in Section~\ref{sec:hydroprofiles}).

In Figure~\ref{pic:09cnd_composition} we present the light curves for models B3, B4, and B5,
which do not satisfactorily fit the observations of PTF09cnd.
They are just through-passage models on the way to constructing a successful model,
but these models illustrate quite well the reasons of choosing chemical composition for the most successful models
shown in Figures~\ref{pic:2zrBobs_09cnd_B0B6} and~\ref{pic:3zrBobs_09cnd}.
The only difference between models B3, B4, and B5 is their chemical composition.
All three models contain about 20~\MS\ in total, and the explosion energy is $4\cdot 10^{51}$~ergs.

We started from the model with C and O in equal proportion (model B4; middle plot).
This is also the standard proportion for the models from our paper \citet{BlSorCOshells_arXiv}.
For this composition and mass, the light curve is too fast in the rising part,
which is quite typical for a CO mixture,
and it is also too blue at maximum.
The problem of short rising time can be solved with taking He instead of a CO mixture
(left plot in Figure~\ref{pic:09cnd_composition}).
In this case, the light curve rises to maximum in a suitable time,
but it still remains too blue.

Changing the proportion between carbon and oxygen in favor of carbon, on the contrary,
removes the blue excess, but preserves the short rising time
(right plot  in Figure~\ref{pic:09cnd_composition}).
Since CO models are more preferable from the observational point of view
(helium lines are not typically observed in SLSN-I spectra),
we have choosen the CO model with an enhanced amount of C as the main model for PTF09cnd.
In order to make the rise time longer, we need to increase the envelope mass
and, therefore, diffusion time.


\subsubsection{The best model}

\begin{figure}

\centering
\includegraphics[width=0.7\linewidth]{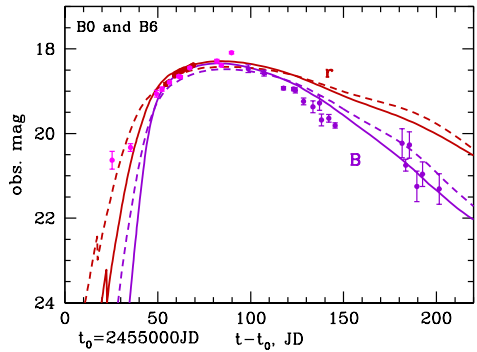}
\caption{\noindent Observed ({\it dots}) and synthetic ({\it lines}) light curves for PTF09cnd in $r$ and $B$ filters.
                   {\it Solid lines} correspond to the model B0 with the outer envelope
                   containing 90\% carbon and 10\% oxygen,
                   {\it dashed lines} correspond to the model B6 similar to B0,
                   but roughly half of C and O in the outer envelope is replaced with He.
        }
\label{pic:2zrBobs_09cnd_B0B6}
\end{figure}

\begin{figure}

\centering
\includegraphics[width=0.7\linewidth]{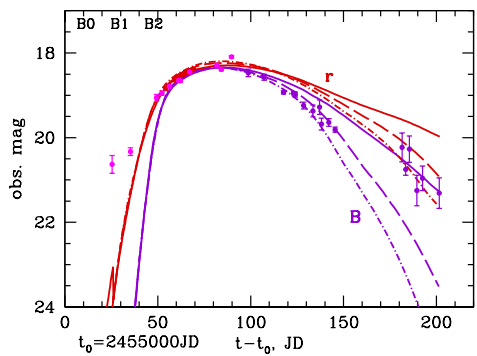}
\caption{\noindent Observed ({\it dots}) and synthetic ({\it lines}) light curves for PTF09cnd in $r$ and $B$ filters.
                   {\it Solid lines} correspond to the model B0 with the static outer envelope,
                   {\it dashed lines} to the model B1 with the expanding envelope
                   with $Ew=0.1$~B as a kinetic energy of the envelope,
                   {\it dash-dotted lines} to the model B2
                   with $Ew=0.3$~B as a kinetic energy of the envelope.
        }
\label{pic:3zrBobs_09cnd}
\end{figure}

The rising time, which more or less fits the observations, was obtained for the model B0
with 49~\MS\ in the envelope and 5~\MS\ in the ejecta of the explosion with the energy of $4\cdot 10^{51}$~ergs.
The C to O ratio for the best model is $9:1$.
The light curves for this model are shown in Figure~\ref{pic:2zrBobs_09cnd_B0B6} with solid lines.
Dashed lines in the plot demonstrate the light curves for the model B6,
which is identical to B0 in the structure, but different in the composition:
its CS envelope contains about 25~\MS\ He and 25~\MS\ CO mixture instead of almost 50~\MS\ of pure CO
(see Section~\ref{sec:presn}).
The changes of the light curve are only minimal:
it becomes a little wider with helium.
This is quite natural, because we replace  half of the  well-absorptive material
with an almost non-absorptive one (see Figure~\ref{fig:opaHe_CO9}),
so the temperature rises more slowly, the gas remains transparent for longer time,
the photosphere shifts outwards more slowly,
and the light curve rises to its maximum also more slowly.

Since the differences of the light curves for these two models are small,
it is difficult to judge which amount of helium is more suitable for observational data
from a photometric point of view only,
unless helium becomes the most abundant element in the CS envelope and gas emission is not dominated
by carbon, oxygen, and heavier elements.
A more detailed study of this topic is required,
which must also include spectral examination,
but this is out of the scope of the current paper.

\subsubsection{Problem with expanding velocity and possible solutions}

{\it Observations vs. modeling.}
Spectral observations of PTF09cnd  show broad features indicating velocities of about $10^4$~km~s$^{-1}$
already near maximum light \citep{QuimbyNatur2011},
while our best model has a static envelope.
In order to test expanding envelopes,
we calculated two additional models, B1 and B2,
which differ from B0 by the kinetic energy of the envelope.

We have already seen in Figure~\ref{pic:10gx_parameters}
that the envelope expansion made the light curve more narrow.
Here the situation is the same.
The width of the light curve for model B0 with a static envelope
(solid line in Figure~\ref{pic:3zrBobs_09cnd})
more or less corresponds to observations,
while the fluxes from models B1 (dashed line) and B2 (dashed-dotted line)
with expanding envelopes decline too fast.
The integral expansion energy of the envelopes is not very high:
$10^{50}$ and $3\cdot10^{50}$~ergs for B1 and B2, respectively.
The velocity is distributed over the envelope so that the inner layers do not move at all,
the velocity grows up toward the outer edge and has a maximum at the outer layers.
The maximal  expanding velocity in the envelope is 750~km~s$^{-1}$ for B1 and 1,300~km~s$^{-1}$ for B2.

Though we did not get a large enough velocity in the current calculations,
and could not fit observations of PTF09cnd with our current expanding models,
this must not be a big problem for the interaction models in principle.
There are several ways to solve the problem.

{\it Rapidly expanding separated shell from pre-explosion.}
The pulsational pair-instability explosion mechanism results in a (multi-)shell structure
with the velocity gradient in the shell larger than we checked with our models
\citep[see, for instance, the velocity distribution for a successful model of SN~2006gy in][]{WooBliHeg2007}.
The use of such a separated shell would solve two problems at once:
the interacting region would remain inside the low velocity gas for a long time,
that would provide a sufficient width of the light curve,
while the broad spectral features would be formed in the outermost high-velocity layers.
The main difference of the latter model from our models B1 and B2 must be in the velocity distribution.
Inner parts of the envelope must have zero or very low velocity up to a rather large radius.
Velocity must  start growing far from the contact discontinuity,
and the velocity gradient must be so high that the outermost layers would expand with the velocity,
which would correspond to the observations, of about 10,000~km~s$^{-1}$.

The outer layers of this model must be well resolved for correct computations,
and time delay effects must be treated in a less primitive way than by {\sc stella}
due to the large radius and high velocities of the outermost layers.
As a challenge for the future, hydrodynamical and radiative transfer calculations
must be combined with detailed modeling of some lines, which show P~Cyg profiles.
The correct computation of ionization and excitation of the rapidly expanding region requires the NLTE approach,
which is beyond the scope of the current paper.
However, we hope to pursue this subject in the future.

{\it Radioactive material.}
Another solution of the problem of getting both wide light curve and wide spectral lines from high-velocity gas is
adding some amount of radioactive material in the SLSN-I model.
If \nifsx\ is generated during the explosion inside an extended envelope,
then the maximum of the light curve can be explained by the interaction,
while the tail is defined by the radioactive decay of \nifsx to \cofsx to \fefsx,
and the width of the light curve does not depend on the expansion of the envelope.

\subsection{Bolometric light curves}
\label{sec:bol}

\begin{figure*}
\centering
\includegraphics[width=0.45\linewidth]{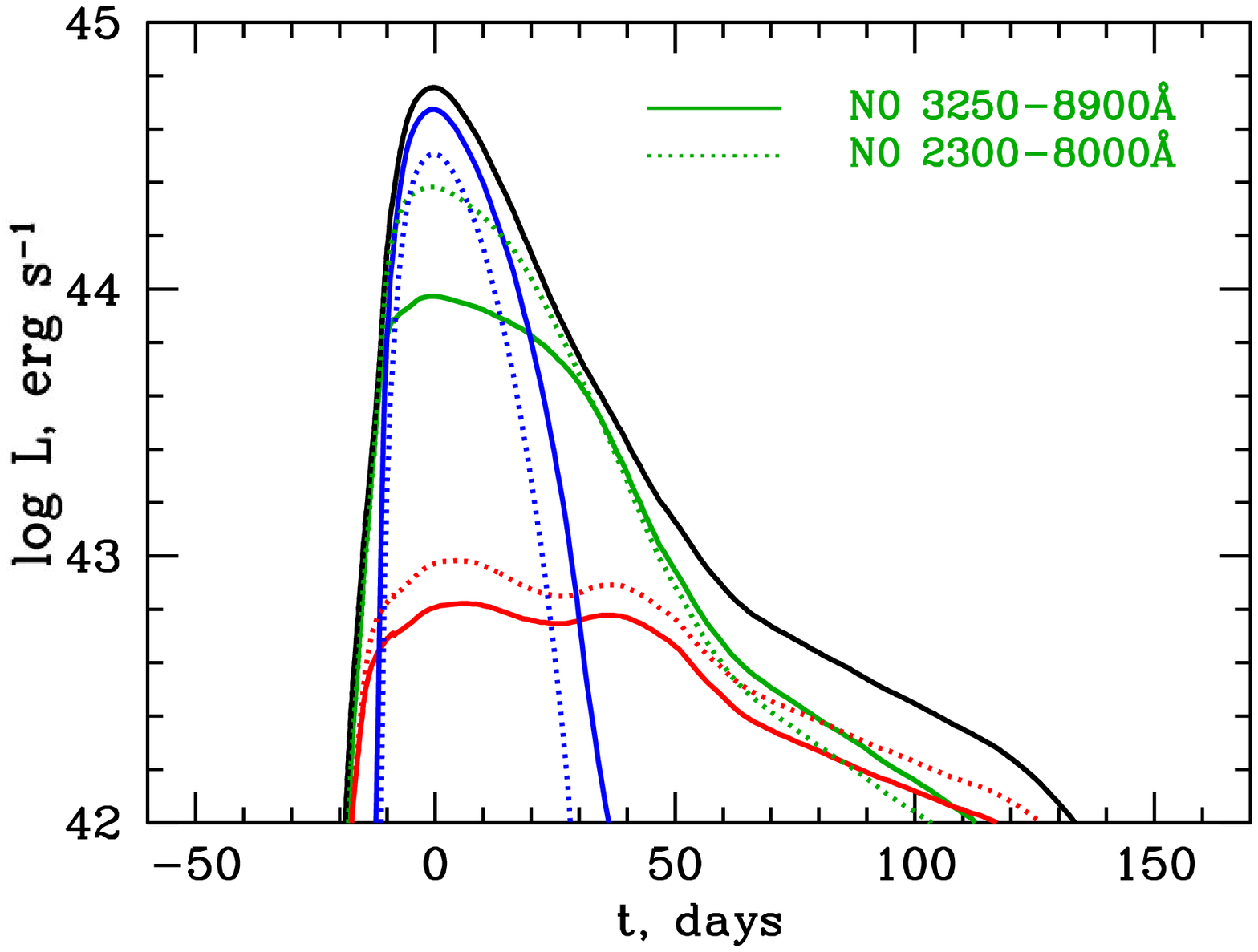} \hfil
   \includegraphics[width=0.45\linewidth]{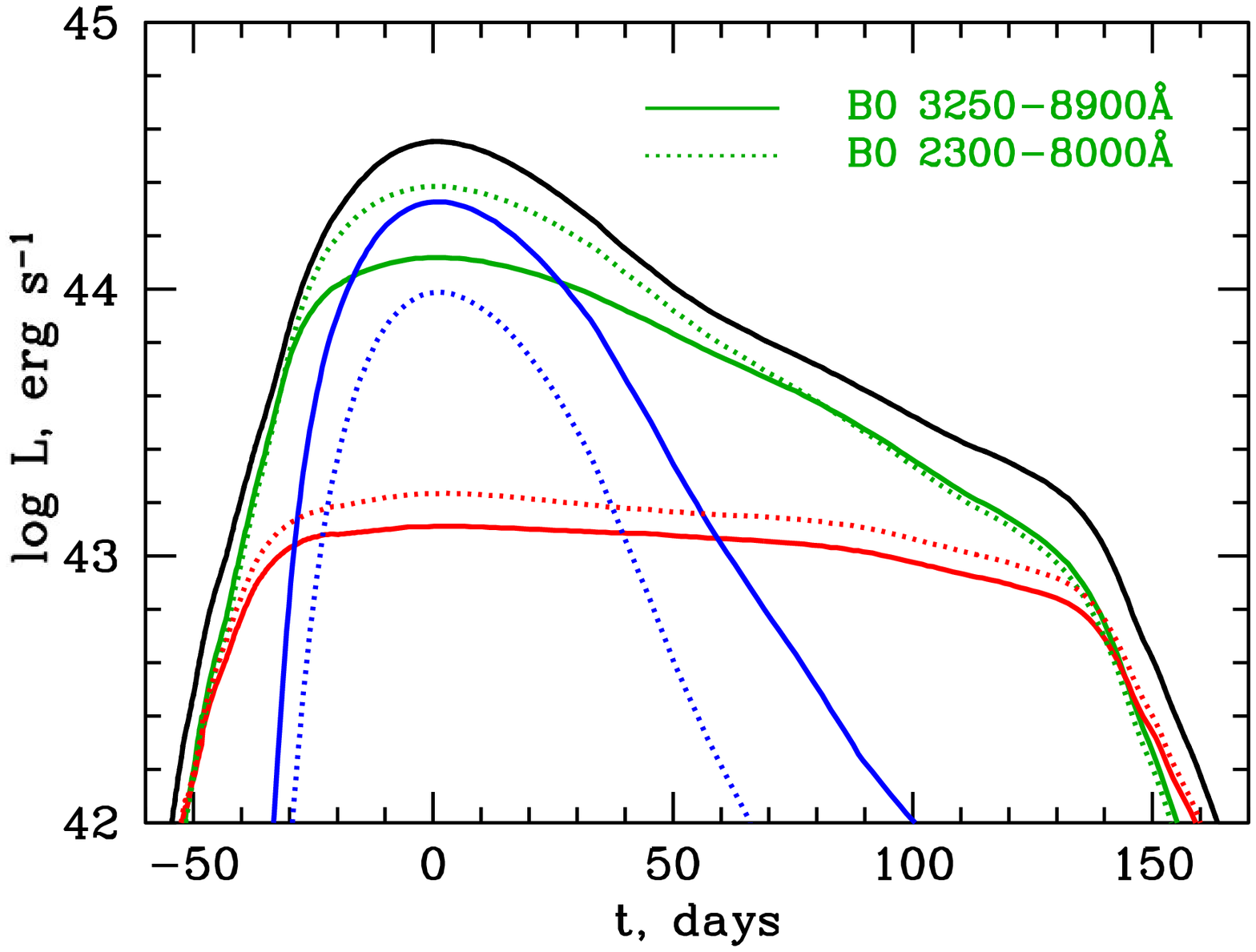}
\caption{Bolometric light curves for the models N0 ({\it left panel}) and B0 ({\it right panel}).
         Black lines are true bolometrics.
         They show the light curves integrated over the whole spectral range used in our calculation
         (1~\AA\ -- 50 000~\AA).
         The black line is a sum of the blue, green, and red lines,
         which represent the parts of the total flux emitted in the extreme ultraviolet,
         visual, and far infrared ranges, respectively,
         but the borders between those ranges are different for the solid and dotted curves.
         The integration ranges for the visual (green) curves are written in the plots.
         The range 3250~\AA\ -- 8900~\AA\ corresponds to the standard {\it UBVRI} range,
         while  2300~\AA\ -- 8000~\AA, to the typical observed rest frame spectral range
         for SLSNe I whose red shift is usually pretty large.
       }
\label{pic:Lbol}
\end{figure*}

Bolometric light curves are easier for theoretical modeling than colored ones,
since less uncertainties are involved in the opacity calculation.
Observers, on the contrary,
must make some theoretical assumptions
to derive a bolometric light curve from their observations,
which typically span only a limited wavelength range.
Therefore, a comparison of observational and theoretical light curves
cannot be  fully consistent:
a given discrepancy might indicate a problem converting the observations to bolometric values
instead of indicating a problem with the theoretical model.
This is why we do not compare the numerically calculated bolometric light curves with the observations,
but demonstrate them alone.

We have integrated the calculated spectra (1~\AA -- 50~000~\AA)
to yield bolometric light curves for the models N0 and B0.
They are shown with black lines in Figure~\ref{pic:Lbol}: left plot is for the model N0, right plot for B0.
With green solid lines, we  plot ``quasi-bolometric'' light curves,
which include integral flux in the wavelength range of {\it UBVRI\-} bands (3250~\AA\ -- 8900~\AA).
This is the range where observational flux can be directly obtained for the local SNe
without applying any additional theoretical assumption about energy distribution in the SN spectrum.

Since all SLSNe observed so far are noticeably red-shifted,
their observations correspond to somewhat bluer rest frame emission.
We have checked the typical rest frame spectral range
and found that SLSN spectra are known roughly in the range of 2300~\AA\ -- 8000~\AA.
The flux for models N0 and B0 integrated in this range is shown with the green dotted lines.
The lines of other colors show the remaining modeled flux, which was not included in the quasi-bolometric range:
the blue  lines show  the integrated flux bluer
than the short wavelength edge of the observable interval discussed above,
while  the red lines show the flux redder than the long wavelength limit.

The figure confirms that the very massive model B0 is much more diffusive around the light curve maximum:
the luminosity in the extremal UV range is very close to the ``visible'' ({\it UBVRI}) value,
especially when we take into account the typical redshift of SLSNe
and shift a part of the UV flux into the observable range
(dotted lines).
In the latter case, the observers are able to measure almost all emission from SLSN.

A large part of the emission from less diffusive SLSNe with narrower light curves still remains
in the UV range near maximum light.
Even when the observed range is shifted to the blue in accordance with the typical SN redshift,
more than  half of the emission still remains in the extreme UV range
and is unobservable from the ground near the light curve maximum.
This means that there is a good chance to observe such SLSNe at even higher redshifts.
Only after about 40 days after the explosion (about 20 days after maximum light),
when the shock wave becomes weaker and the temperature within the envelope-ejecta system falls down
(see Figure~\ref{fig:hydroN0}),
does the visible wavelength range carry most of the emitted radiation.

\begin{figure}
\centering
\includegraphics[width=0.7\linewidth]{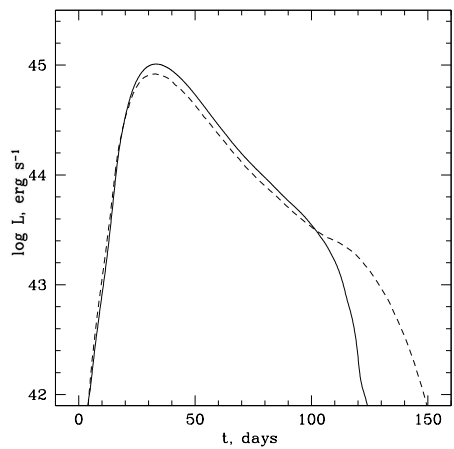}
\caption{Bolometric light curves for one of the models constructed to reproduce the PTF09cnd light curve.
The solid and dashed lines represent, respectively,
the {\sc rada} and {\sc stella} calculations in the observer's frame of reference.
       }
\label{pic:LbolStellaRada}
\end{figure}

An estimation of the Eddington factors is rather crude in {\sc stella},
and time delay effects are not taken into account.
To check whether the radiation transfer and time delay effect could affect the light curve,
we performed a simulation for one of the models for PTF09cnd  using the {\sc rada} code
\citep{Tolstov2003,Tolstov2010}.
In comparison to {\sc stella}, {\sc rada} calculates exact Eddington factors at each time step
and takes into account time delay effects more accurately.
The bolometric light curve (Figure~\ref{pic:LbolStellaRada}) does not reveal significant deviations
from {\sc stella} results up to 100~days after an explosion,
until the system becomes more transparent and the shock becomes directly visible.
Deviations grow only when the SLSN luminosity weakens by 1.5 orders of magnitude,
a phase when typical SLSNe become almost unobservable.
This means that more accurate modeling of the radiation transfer is not so important for similar models
at least near the stage of the maximum light,
while the envelope is essentially optically thick.

\subsection{Spectra}
\label{sec:spectra}

\begin{figure*}
\centering
\includegraphics[width=0.45\linewidth]{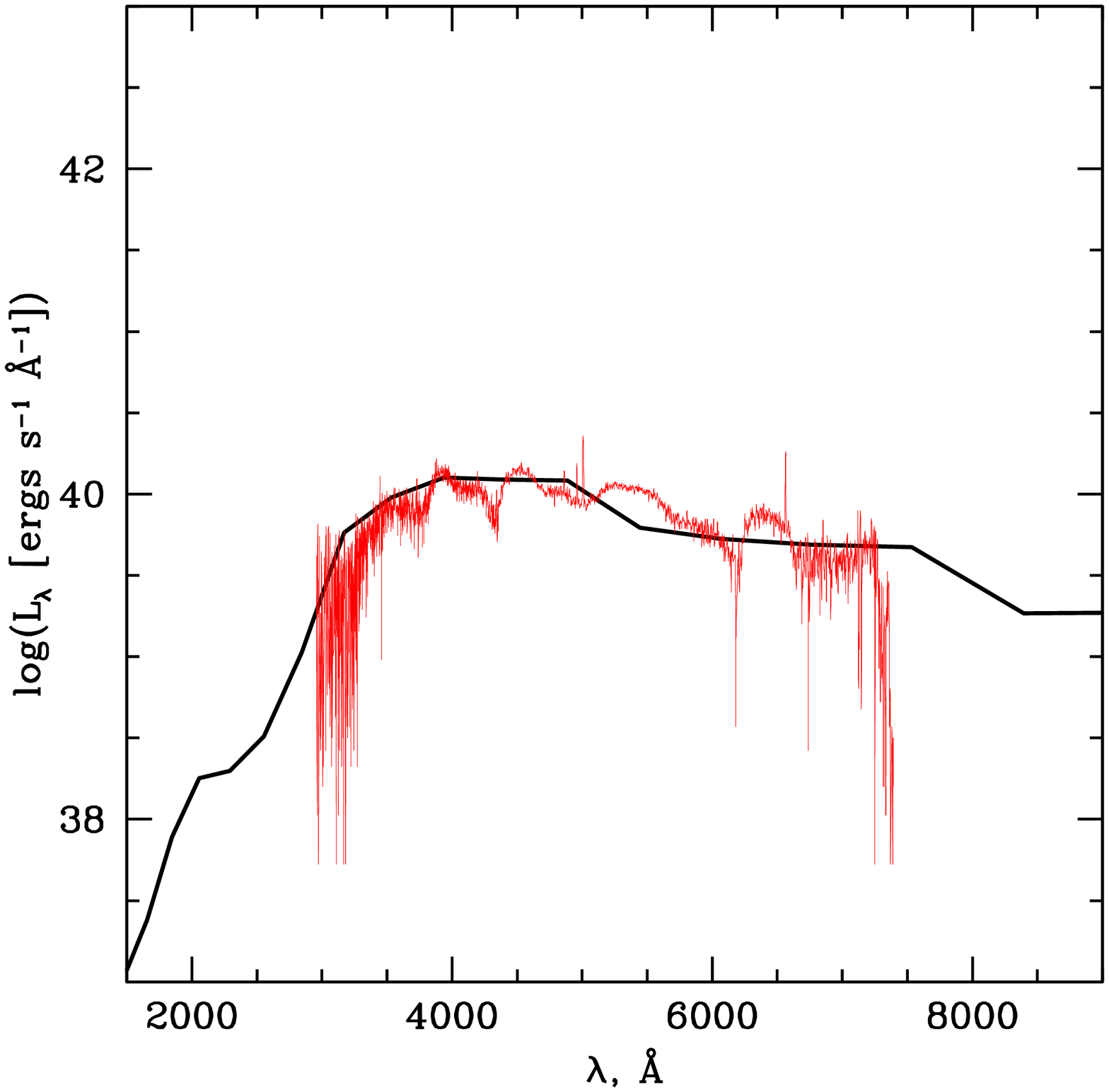}\hfil
\includegraphics[width=0.45\linewidth]{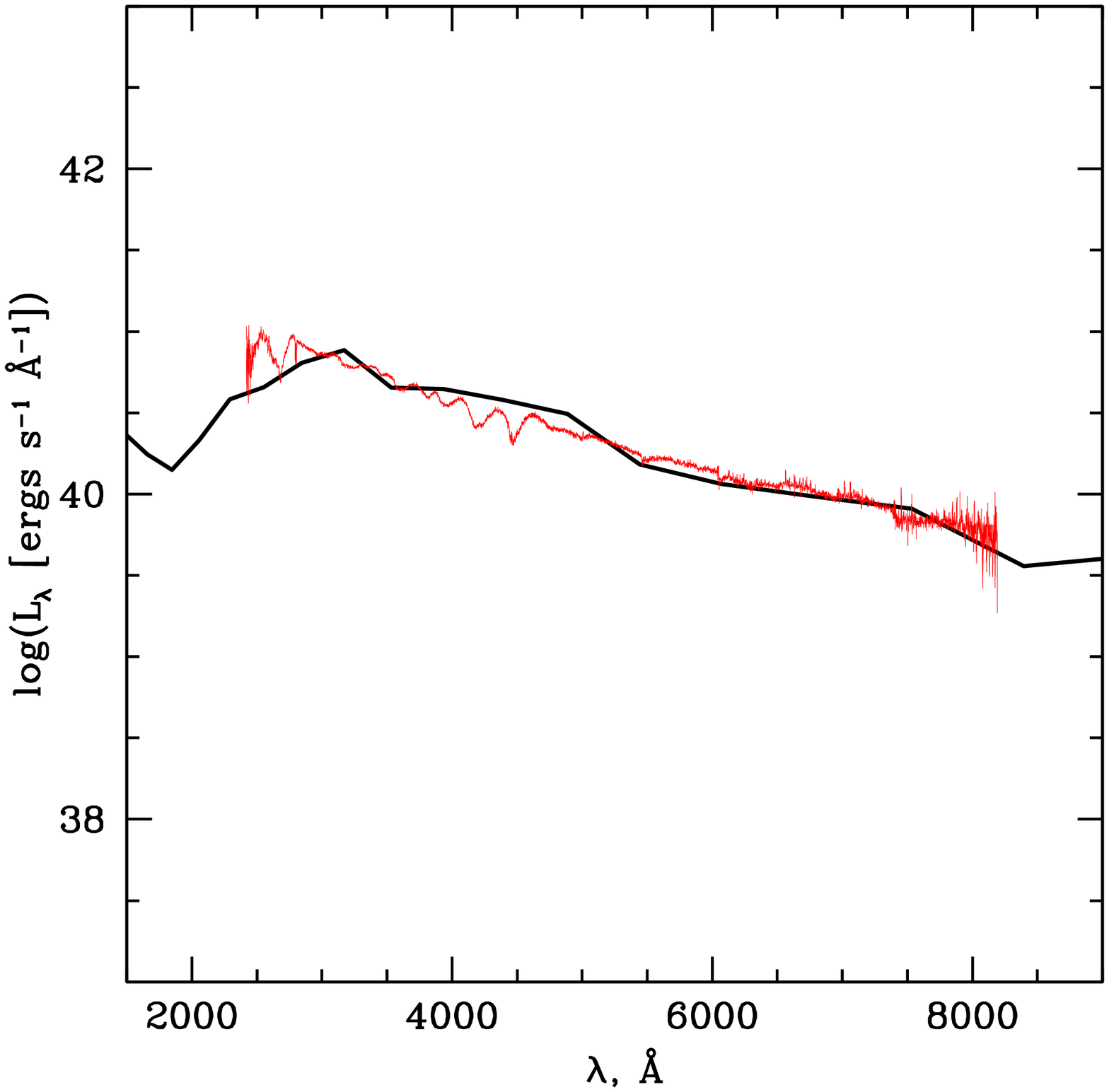}
\caption{\noindent Rest frame observed ({\it red}) and modeled ({\it black}) spectra.
   The {\it left panel} compares the observed spectrum of SN~2010gx at day $+27$ \citep{Quimby2013}
   with that of model N0 at day $+32$ after the maximum in {\it B}-band.
   The {\it right panel} compares the observed spectrum of PTF09cnd at day $-20$ \citep{Quimby2013}
   with that of model B0 at day $-20$.
   The observed luminosities are in arbitrary units and can be shifted along the {\it y-}axis
   for better fitting to the model.
        }
\label{fig:sp10gx_09cnd}
\end{figure*}

\begin{figure*}
\centering
\includegraphics[width=0.45\linewidth]{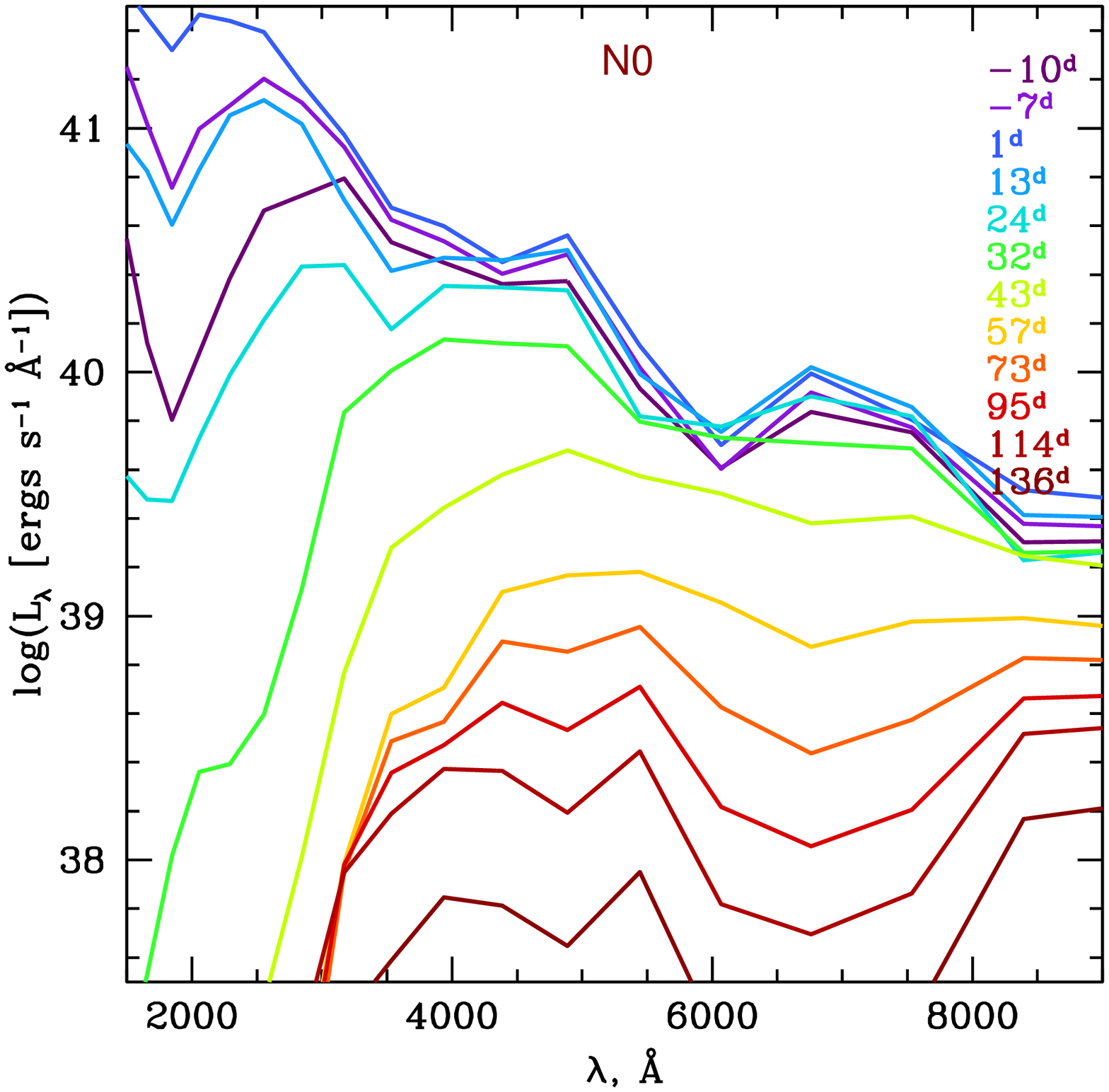}\hfil
\includegraphics[width=0.45\linewidth]{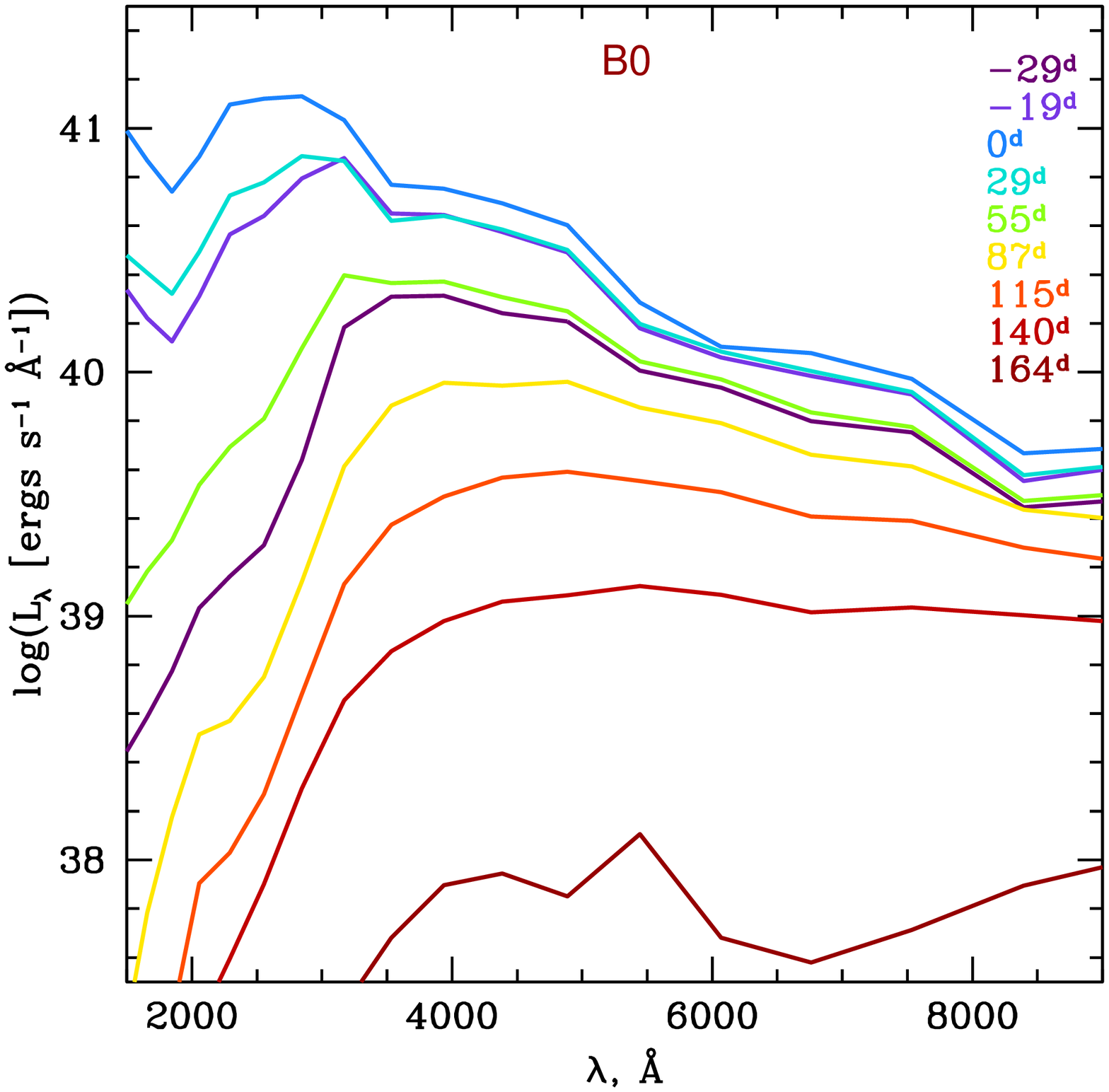}
\caption{\noindent Time evolution of the rest frame spectra for the models N0 ({\it left}) and B0 ({\it right}).
   The color of each spectrum corresponds to that of the age written in the upper right corner of each panel.
\label{fig:animspN0B0}
        }
\end{figure*}

We have shown in Sections~\ref{sec:res10gx} and \ref{sec:09cnd}
that the broad band light curves of SN~2010gx and PTF09cnd are all in good agreement
with the interacting model with strongly different CSM mass.
An even more difficult problem is to get a good fit to the SLSN spectra.
{\sc stella} typically uses a grid of only 100 frequency bins,
so we are not able to reproduce all spectral details.
Lines are all averaged within the bins of some hundred Angstr\"om width.
So we can compare the observed spectra only
with {\sc stella}'s crude spectral-energy distribution.
Figure~\ref{fig:sp10gx_09cnd} provides this comparison.
Time evolution of the spectra for our best models N0 and B0 are shown in Figure~\ref{fig:animspN0B0}.

Along with the fast photometric evolution of SN 2010gx comes relatively fast spectroscopic evolution.
We compare a spectrum of SN~2010gx from \citet{Pastorello2010} with the age estimated as $+27$ days after maximum
with our modeled spectrum in Figure~\ref{fig:sp10gx_09cnd}.
We find the best fit to this spectrum for model N0 at $+32$ days after the maximum
in the {\it B} band.
At $+27$ days, the model spectrum was a bit too hot:
the UV emission exceeds the observations.
Nevertheless, we find the 5-day phase difference to be negligible,
especially when including the potential error in the observed date of maximum.
Thus, we consider the model and observed spectra to be in good agreement.

Our model calculation shows that the cut-off wavelength of the UV emission
shifts rather fast during the evolution of the object after the maximum light (Figure~\ref{fig:animspN0B0}).
Comparison of a few observed and modeled spectra within a couple of weeks during the fading phase would be very useful
for examining a model in the future.

PTF09cnd has a wide light curve and the observers were able to get its spectrum well before
the maximum.
The spectrum shown in the right panel of Figure~\ref{fig:sp10gx_09cnd} was taken
20 days before maximum \citep{QuimbyNatur2011}.
At this stage, the emitting gas is much hotter than that from the fading stage on the left panel
and the drop-off wavelength is still out of the measured range.
However, the slopes of the observed and model B0 spectra at the same age,
$-20$ days, coincide well.

The good agreement in observed and modeled spectra supports the ejecta-CSM interaction
as an explanation for the wide range of non-hydrogen SLSNe.

\vfill

\section{SUMMARY AND DISCUSSIONS}
\label{sec:summary}

\subsection{Main results}
\label{subsec:res}

The aim of the current work is to test the role of shock
interaction in producing light in the non-hydrogen CS envelopes around SLSNe-I
and to see if the bulk of SLSNe-I can originate from this kind of system.
Our calculations indicate that if this envelope structure forms in nature,
then it can indeed give rise to extremely bright and long-lived supernova light curves.
There have been no detailed numerical simulations of the broad band light curves of SLSNe-I
except for
 \citet{BlSorCOshells_arXiv} and \citet{BSB12dam}.
In the current paper, we succeeded in reproducing the narrow light curves of SLSNe-I as well as the broad ones.
We got the correct duration of the light curves and correct fluxes in different bands.
We also get very good agreement in reproducing continuum spectral-energy distributions for both cases.
We show that the details of the light curves over the fading stage
depend on the structure of the CS envelope.

Both SN~2010gx and PTF09cnd can be
explained with moderate explosion energies
$\sim (2 - 4)\cdot 10^{51}$~ergs for an envelope extending up to $10^{16}$ cm with an almost flat density profile.
Energetics is one of the main advantages of the interacting mechanism of SLSNe
compared to pair-instability or magnetar-powered SLSNe:
the shock wave is very effective in producing thermal energy,
so models with CSM require explosion energies only a little higher than that of ``normal'' SNe,
while in all other mechanisms the energetics must be at least an order of magnitude higher than normal.

The mass of the surrounding envelope must be high enough
for the effective transformation of kinetic energy of the ejecta
into observed light.
The main difference between models with narrow and broad light curves
is their masses.
Models for SN~2010gx require 5 to 10~\MS\ of carbon and oxygen in the envelope,
which is right in line with the ejected masses expected for pre-explosion
from the pulsational pair instability  mechanism,
though the possibility of loosing all hydrogen and even helium long before the explosion is controversial.
A much more massive C+O envelope with $\sim$ 55~\MS\ is needed
to reproduce the broad band light curves of PTF09cnd.
There are predictions, however, for stars in the last stages of stellar evolution
with the requisite large helium and carbon-oxygen core masses
\citep{Waldman08StarEvol,OhkuboNom_mergers}.

Models with an extended envelope contiguous to the ejecta fit SN~2010gx and PTF09cnd better,
but a detached shell or another feature in the envelope's density distribution might also be a good possibility
to  reproduce features on the light curve for some other SLSNe.
Models with detached shells can change the light curve shape during the fading stage.

\subsection{Limitations}
\label{subsec:limits}

We encounter two kinds of problems in simulating SLSNe with the shock interaction mechanism,
physical and numerical,
which limit either the applicability of the model to the observed SLSNe
or the applicability of the numerical code to the problem.

\vfill

\subsubsection{Physical problems}

{\it Mass of hydrogen-poor envelope.}
One of the obvious physical limitations of the interaction models is provided
by the duration of the light curve tail.
Broad light curves require non-hydrogen envelopes of very large masses in the shock interacting model to explain SLSNe.
For example, the model B0 for PTF09cnd contains almost 50~\MS\ of C and O in the extended envelope.
If SLSNe-I with longer light curve tails are discovered,
the interaction scenario would require even higher envelope masses,
which may not be very realistic.
The solution of the problem may require including an appreciable amount of radioactive material
to the shock interaction model.
Pure pair instability SN models are too diffusive and have a rise time to maximum that is too long
\citep{NichollMagnetar}, though the amount of \nifsx\ they contain can easily explain the SLSNe-I tails.
The shock interaction models can help to solve the problem of rising time \citep{BSB12dam}.
Thus, the combined models of shock plus radioactivity may provide the best solution for the problem.
Then the shock will be responsible for the maximum of the light curve,
while the radioactivity will determine the behavior of the tail.
This may be most relevant to SLSNe with long tails of constant decline.

{\it High velocities.}
The combined model (radioactivity plus interaction) can also help to explain
the high velocities observed in SLSNe-I.
Another type of shock interaction supernovae, SNe~IIn,
clearly demonstrates narrow emission lines produced in the shells.
Not so in SLSNe-I: narrow circumstellar lines are not seen \citep{Pastorello2010,QuimbyNatur2011}.
There is no easily excited hydrogen in this type of SNe, and the most abundant elements
(probably, carbon and oxygen) should  present  as C~II and O~II ions in the envelope.
These ions do not have many strong lines in visible light.
It is not easy to identify C and O lines in the photospheric stage in SNe~I \citep{Young2010},
and for SLSNe they should be excited even in the absence of
the radioactive material.
Thus, one has to look for weak lines in noisy spectra.
This problem certainly deserves further investigation to account
for different conditions of ionization/excitation of shells under the shock radiation.
Shock interaction models with fast expansion of the envelope produce rather narrow light curves.
For the SLSNe-I with the broadest light curves we again encounter the problem that,
under the assumption of an expanding CS envelope,
extremely high envelope masses are required to fit the light curves.

\subsubsection{Numerical problems}

Computation of the shock-interacting models on a fine grid is often time consuming.
In some cases, semianalytic models \citep[e.g.,][]{Chatz12,Chatz13} can help,
but  accurate calibration against more sophisticated numerical calculations may be needed.
Currently, we see discrepancies in some parameters of our models with the models from \citet{Chatz13}
for the same SLSNe.

The fits to SLSN fluxes in individual filters are not yet perfect in our simulations.
This is natural: we have taken quite arbitrary and primitive
chemical compositions, density distributions, etc.
However, there is no principal problem for us to reproduce typical brightness and duration of the light curves
within pure interacting models.
One can try to build a better fit to the observations by varying the initial
conditions in the model, especially, the density distribution in the envelope and its chemical composition.
However, it seems that it is too early to optimize the models along these lines.
This optimization will probably not give us a true insight and
a better understanding of the problem.

{\it Expansion opacities.}
Before optimizing the models, we have to improve the physics in our simulations.
One of the main difficulties is  the treatment of line opacity.
Sophisticated Monte-Carlo codes
\citep[e.g.,][]{Kasen2006,Kasen2007,Sim2007,Kromer2009} or
direct integrations \citep{Dessart2009} for radiative transfer in spectral lines are not
applicable for flows with complicated non-monotonic velocity profiles.

In our simulations, we use the approximation of expansion opacity
\citep{Bli1996,Bli1997}.
Some modifications are needed for
corrections  of the expansion effect not only in the flux equation,
but also in the energy equation as discussed by
\cite{SorBli2002}.
Moreover, there is another complication with the anisotropic velocity
gradient.

{\it Velocity gradient in expansion opacity calculations.}
In the standard {\sc stella} setup, the expansion opacity in lines is treated as for type~Ia supernovae,
where expansion is homologous and  velocity gradient is isotropic, $dv/dr=1/t$,
with $t$ -- time elapsed after the explosion.
In the interacting models of SLSNe, the source of light is a long-living shock wave.
In this case, the velocity gradient is not isotropic
and changes along the radius.
Moreover, it is negative on the shock front, with  $|dv/dr| \gg 1/t$ during several months,
while the shock passes through the envelope.
The influence of the expansion on the opacity in this region must be stronger than in the isotropic case.
The current version of {\sc stella} does not take into account this anisotropy in the expansion opacity calculation,
and this is a source of uncertainty in resulting light curves.
To estimate the uncertainty of line opacity calculations, we have run several tests when
the expansion opacity is calculated with the fixed value $dv/dr=1\,\mbox{day}^{-1}$,
which crudely takes into account the fact that the light originates from the shock wave
with the enhanced value of the velocity gradient \citep{BlSorCOshells_arXiv}.
The observed flux becomes higher in many bands, and the shape of
light curves changes noticeably as compared with the standard {\sc stella} calculations,
which shows the uncertainty range of our results.

{\it Dimensionality.}
The main complication to the whole picture is possible fragmentation of the dense shell.
The attempts on multi-D treatment of SN ejecta evolution are rather old
\citep{TenT1991,CheBlo1995,BloLunChe1996}, more recent
results and references may be found in \citet{Dwar2007,Dwar2008}.
See also \citet{VMarle2010} for the case of SN~2006gy, but without real treatment of radiative
transfer.
There are several 3D MC transport codes
\citep{Hoeflich2002,Lucy2005,Kasen2006,Kasen2007,Sim2007,Tanaka2008,Kromer2009},
but they are not actually coupled to hydrodynamics and there are many
difficulties in doing this \citep{Almgren2010}.

{\it NLTE.}
Full NLTE treatment is needed to predict spectra,
but very little is done on this even for SN~IIn.
For example, \citet{Dessart2009} are successful in reproducing the spectra
of SN~1994W in a set of atmospheric models,
though their method is applicable only to monotonic velocity structures,
not to shocked shells.
Moreover, one should be cautioned about the relation of the ``photospheric''
radius found by \citet{Dessart2009}, which shrinks,
and the radius of the shocked shell in SNe~IIn, which grows.
This has already been explained by \citet{SmithChor2008,Smith2010}.

\subsection{Comparison with the analytical model}
\label{subsec:analyt}

\begin{figure}
\centering
\includegraphics[width=0.8\linewidth]{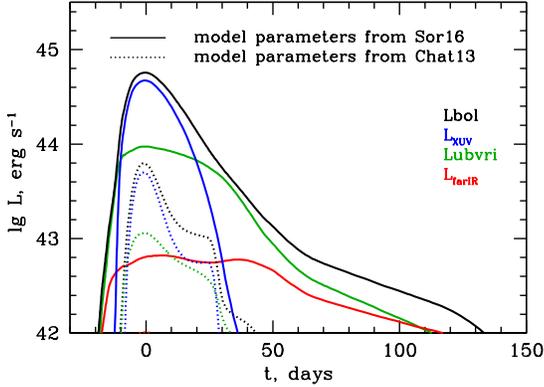}
\caption{Bolometric light curves for model N0 (the best one for SN~2010gx among our models; {\it solid lines})
and for results of numerical run with the parameters from analytical model by \citet{Chatz13} ({\it dotted lines}).
Colors mean the same as in Figure~\ref{pic:Lbol}. Green lines are the quasi-bolometric ones integrated over the wavelength
range 3250~\AA\ -- 8900~\AA.
       }
\label{pic:analyt}
\end{figure}

In our numerical modeling, we take into account a lot of physical
processes and examine their effects on the light curves.  To
demonstrate that this would be a challenging task in the analytical
approach, we calculated the light curves for SN~2010gx by adopting the
same model parameters as in the analytical model by \citet{Chatz13}
for SN~2010gx in the interaction scenario: $M_{\rm ej}=9.7M_{\textstyle \odot}$, $R_{\rm ej}=2\cdot 10^{13}$~cm,
$M_{\rm CSM}=1.64M_{\textstyle \odot}$, $M_{\rm Ni}=0$, $E_{\rm SN}=1.14\cdot 10^{51}$~ergs,
$R_{\rm CSM}=1.58\cdot 10^{15}$~cm.
\citet{Chatz13} provided the results for two models of the interaction scenario with a different density distribution
in the CS envelope.
We chose the model with $\rho=r^{-2}$ because it is more similar to the density distribution in our models.
The results of our numerical run of the model with the parameters listed above are shown
by the dashed lines (Chat13) in Figure~\ref{pic:analyt}.  For comparison,
the light curves for our model N0,
which are in good agreement with broad-band fluxes from SN~2010gx, are shown by the solid
curves (Sor16).

Our numerical light curves Chat13 are noticeably different from the
analytic model by \citet{Chatz13}.  The light curves Chat13 are less luminous and much
shorter than the analytical formula predicts (Figure~4 of \citep{Chatz13}).
Most probably, the difference in brightness is explained by an order of magnitude smaller radius of the SC envelope
in Chat13 model, while the difference in the duration is due to a smaller diffusion time
as a consequence of a smaller CS envelope mass.
Such a difference in light curves is
very similar to that found in the comparison between the analytic
and numerical models for Type II SLSN 2006gy (Figure~14 of \citep{Moriya13_09gy}).
Further details will be presented in the
forthcoming paper.


\subsection{Open questions}
\label{subsec:openQ}

\subsubsection{Helium}

Our best models for SN~2010gx and PTF09cnd require circumstellar envelopes of 5 to 50~\MS, which consist of
70--90\% carbon and 30--10\% oxygen and must be formed before the explosion.
Stellar evolution theory tells us that it is difficult to create a star
with almost all helium lost well before the SN explosion
and such a huge amount of almost pure carbon and oxygen lost a few months before the explosion.
Our modeling excludes the possibility of a purely helium CS envelope
because it demonstrates wrong colors near the maximum.
In order to make a more evolutionary plausible model,
we implemented one chemically non-uniform model for PTF09cnd:
the CS envelope of model B6 consists of 25~\MS\ helium, 22~\MS\ carbon, and 3~\MS\ oxygen,
while the ejecta is mostly CO.
This model confirmed our prediction
that unless helium is overwhelmingly abundant,
the calculated light curve is changed only slightly.
A larger amount of helium may reduce the envelope mass needed to reproduce
the slow rise time of the SLSN, because the rise time
is longer for a pure helium envelope of the same mass.
On the other hand, the pure helium envelope makes the SN model bluer than the observed SN near maximum,
as seen in model B3 for PTF09cnd
(Figure~\ref{pic:09cnd_composition}).
Thus, more detailed work related to the dependence of the light curve shape on the chemical composition is still necessary.

Until recently, there was no sign of helium in SLSN-I spectra,
but the observations of \citet{YanQuimby13ehe} changed the situation.
Not only helium, but even hydrogen is observed in the late (about a year after the explosion) specta of iPTF13ehe
and some other SLSNe-I, which means that the CS envelope in the interacting scenario must be non-homogeneous,
and its outermost layers contain He and H.
Such systems are easier to understand from the point of view of stellar evolution.
This clearly demonstrates that  chemically uniform models from our current work are a bit idealized.
In the future we plan to investigate more complicated kinds of models with stratified CO--He--H layers
as well as models with different mixtures of CO$+$He.
Both of them look more natural than the current models,
which are still very good as a first approach to numerical modeling of SLSNe-I in the interaction scenario.

Our first approach to the mixed CO$+$He models shows
that the light curve is not much affected
unless the amount of helium would become sufficiently large,
because the rising time must be determined by a more opaque element.
Only colors are changed a little.
Here we agree with the conclusion of \citet{piro14}
that the total helium mass of stripped envelope SNe is difficult to be measured
from simple light curve modeling.
Future research must answer two main question regarding the mixed models:
put more definite upper limit to the mass fraction of helium in the CS envelopes
and estimate how much CS envelope mass can be reduced for the most abundant He models.
The stratified models can possibly provide us with some interesting features on the light curve
when the photospheric radius would stumble upon the borders between different element layers.

Spectroscopically, models with and without helium can also look very similar.
For example, \citet{dessart15} argue that He features can easily be hidden in the spectra.
They described a model for type Ic supernova
with  30\%  surface helium mass fraction, and found that He I lines were
always absent in the visible light. There is no source of the excitation
of helium if radioactive material is not mixed toward the surface.
Only a weak 10 830~\AA\ line is present.
In our models, we normally have a rather low color temperature
(see Figure~\ref{fig:animspN0B0}). For early times, it may reach $2\cdot 10^4$~K, so some
thermal excitation  of He may be expected as in B-stars, but the
presence of large amounts of C may hide it, like in \citet{dessart15}.
This question deserves further investigation, which is outside the
scope of the current paper.

Direct evidence of the existence of hydrogen- and helium-poor envelopes around an exploding star has been seen,
for example, by~\citet{BenAmi2014}.
They present the observations of Type Ic SN2010mb lacking
spectroscopic signatures of H and He (which
should be the case for the models that we build in the current work).
SN2010mb has a slowly declining light curve ($\sim600\,$days)
that cannot be powered by $^{56}$Ni/$^{56}$Co radioactivity.
The signatures of interaction with hydrogen-free CSM include a blue
quasi-continuum
as well as narrow oxygen emission lines that require high densities
($\sim10^9$cm$^{-3}$).
This is again similar to our models of the dense CS envelope.
The difference is in the mass involved in the interaction:
they estimate that the total amount of interacting gas is about 3~\MS,
while we need 50~\MS\ for our most massive model.
The estimations of the parameters of the H- and He-poor envelope around SN~2010mb by~\citet{BenAmi2014}
are in agreement with the masses and compositions of some models of pulsational pair-instability SNe
\citep{Yoshida16}.

\subsubsection{Progenitor}

The main question, which remains unclear, is the origin of the extended and dense CS envelope,
either carbon-oxygen, helium, or a mixture of both.
What is the time scale of the formation of the envelope?
How far can the  envelope extend?
What is the density profile and the temperature of matter before the core explosion?
These questions deserve further investigation.

The theory of stellar evolution suggests several possible scenarios,
which can result in such a system with a CS envelope.
One of them is an explosion through the  pulsational pair-instability mechanism as a means
to form a detached CS shell.
For instance, in the models of \cite{WooHeg2015} ejected masses and  energies
of preexplosions for the stars with helium core masses of about 50~\MS\
are similar to those needed to form the CS envelope
with the parameters we use in our calculations.
Evolutionary calculations of  helium cores of the most massive stars in \citet{OhkuboNom_mergers}
also lead to the loss of several tens of solar masses of the gas before the final explosion.
A major part of the lost material is helium and oxygen in their calculation.
The fact that SLSNe-I are typically observed in low-metallicity, star-forming galaxies
favors the explosion of stars with very high initial masses,
which are able to retain a large part of their mass during the evolution
and loose a large amount of mass just before the explosion.

Evolutionary calculations of very massive stars (a few hundreds of solar masses) of different metallicities
show that most of the hydrogen burns into helium during the main sequence evolution of such stars,
before the stage of an intensive mass loss \citep{YusofHirschi2013}.
This happens due to convective cores of such stars that extend up to outer layers of stars.
In the very later phase of evolution, helium shell burning produces
a substantial amount of carbon (more than oxygen) in the helium shell.
In addition, the rotation of a star forms a meridional circulation,
which also helps most of the hydrogen burn into helium
(and, possibly, some part of helium into C$+$O) even at the surface.
The combination of the formation of such a hydrogen-poor star
with a powerful blowout from it as a wind in the latest stages of evolution,
or with the ejection due to the pulsational pair-instability,
can lead to the formation of a system required in this paper as an initial model for SLSN-I.
The pulsational pair-instability mechanism is more preferable
because the explosive processes can more easily produce very high velocities of the CS envelope,
of the order of 10,000~km/s, observed in the SLSN-I spectra.
These velocities are too high to have originated from a wind.

One can also speculate about the merger (or even multiple mergers) of a CO core from an evolved WR star
with another hydrogen-deficient star,
like a neutron star, a white dwarf, or another WR star
\citep{FryerBenzHer,TaamSand,GlebGabdeMinkMultMerg,BarkovKomiss,Chevalier12}.
A rich variety of probable binary pre-SLSN evolution is considered in \citet{vdHeuvel2013}.
These events are very probable in close binary systems and in the central regions of young stellar clusters
(like the ejection of the common envelope in pre-SN~Ia binaries considered by \citep{IbenTut1984}).
The merging can lead to a moderate explosion with energy of a few percent of
$10^{51}$~ergs which forms a relatively large  cloud of matter around the combined core.
The size of this envelope can be enough to provide a long-living shock wave
when the central core collapses  within a few years after the merger.
Such an extended envelope would produce a very bright supernova,
though the envelope expansion velocity would doubtfully be high enough.
More detailed numerical calculations of merging systems are needed.

Analyzing all the possibilities above, we cannot choose one of them with certainty.
However, with the current state of knowledge, the most plausible progenitor of SLSNe-I is a very massive star
(maybe with fast rotation), in which all of the hydrogen and a large part of the helium is burned out or
expelled (or lost) during presupernova evolution
and which is still massive enough to pass through the pulsational pair-instability stage.
This scenario can explain both chemical composition and high envelope velocities in the interaction model for SLSN.
We conclude that, provided the formation of rather dense and extended
circumstellar envelopes by any pre-SN scenario, extremely powerful events,
like SN~2010gx and PTF09cnd, can be explained with moderate explosion energies
without invoking any radioactive material.
In principle, this mechanism can be the same for all SLSNe-I observed so far.

\acknowledgments

SB is thankful to Victor Utrobin and to Takashi Moriya for discussions on bright SN~I
spectra, light curves, and on numerical modeling of radiation dominated shocks with our code
{\sc stella}.
KN and AT would like to thank Raphael Hirschi for a discussion
on the evolution of very massive stars and the formation of dense CSM
during his stay at Kavli IPMU as an affiliate member.

The work in Russia (calculation of the SLSN light curves) was supported
by a grant of Russian Science Foundation 14-12-00203.
The work in Japan has been supported by the World Premier
International Research Center Initiative (WPI), MEXT, Japan, and by
the Grants-in-Aid for Scientific Research of the JSPS (26400222 and 16H02168).

\end{document}